\documentclass[twocolumn,showpacs,amsmath,amssymb,pra,superscriptaddress]{revtex4-1}

\usepackage{graphicx}
\usepackage{float}
\usepackage{dcolumn}
\expandafter\let\csname equation*\endcsname\relax
\expandafter\let\csname endequation*\endcsname\relax
\usepackage{amsmath}
\newcommand{\ket}[1]{|#1\rangle}
\newcommand \be{\begin{equation}}
\newcommand \ee{\end{equation}}
\newcommand \bea{\begin{eqnarray}}
\newcommand \eea{\end{eqnarray}}
\newcommand \bse{\begin{subequations}}
\newcommand \ese{\end{subequations}}

\begin{document}

\title{Application of adiabatic passage in Rydberg atomic ensembles for quantum information processing}

\author{I.~I.~Beterov}
\email{beterov@isp.nsc.ru}
\affiliation{Rzhanov Institute of Semiconductor Physics SB RAS, 630090 Novosibirsk, Russia}
\affiliation{Novosibirsk State University, 630090 Novosibirsk, Russia}
\affiliation{Novosibirsk State Technical University, 630073 Novosibirsk, Russia}

\author{D.~B.~Tretyakov}
\affiliation{Rzhanov Institute of Semiconductor Physics SB RAS, 630090 Novosibirsk, Russia}
\affiliation{Novosibirsk State University, 630090 Novosibirsk, Russia}

\author{V.~M.~Entin}
\affiliation{Rzhanov Institute of Semiconductor Physics SB RAS, 630090 Novosibirsk, Russia}
\affiliation{Novosibirsk State University, 630090 Novosibirsk, Russia}

\author{E.~A.~Yakshina}
\affiliation{Rzhanov Institute of Semiconductor Physics SB RAS, 630090 Novosibirsk, Russia}
\affiliation{Novosibirsk State University, 630090 Novosibirsk, Russia}

\author{I.~I.~Ryabtsev}
\affiliation{Rzhanov Institute of Semiconductor Physics SB RAS, 630090 Novosibirsk, Russia}
\affiliation{Novosibirsk State University, 630090 Novosibirsk, Russia}

\author{M.~Saffman}
\affiliation{Department of Physics, University of Wisconsin, Madison, Wisconsin, 53706, USA}

\author{S.~Bergamini}
\affiliation{The Open University, Walton Hall, MK7 6AA, Milton Keynes, UK}

\begin{abstract}
We review methods for coherently controlling Rydberg quantum states of atomic ensembles using Adiabatic Rapid Passage and Stimulated Raman Adiabatic Passage. These methods are commonly used for population inversion in simple two-level and three-level systems. We show that adiabatic techniques allow us to control population and phase dynamics of complex entangled states of mesoscopic atomic ensembles for quantum information processing with Rydberg atoms. We also propose several schemes of single-qubit and two-qubit gates based on adiabatic passage, Rydberg blockade and F\"{o}rster resonances in Rydberg atoms. 
\end{abstract}

\maketitle

\section{Introduction}

\begin{figure}[!t]
\center
\includegraphics[width=\columnwidth]{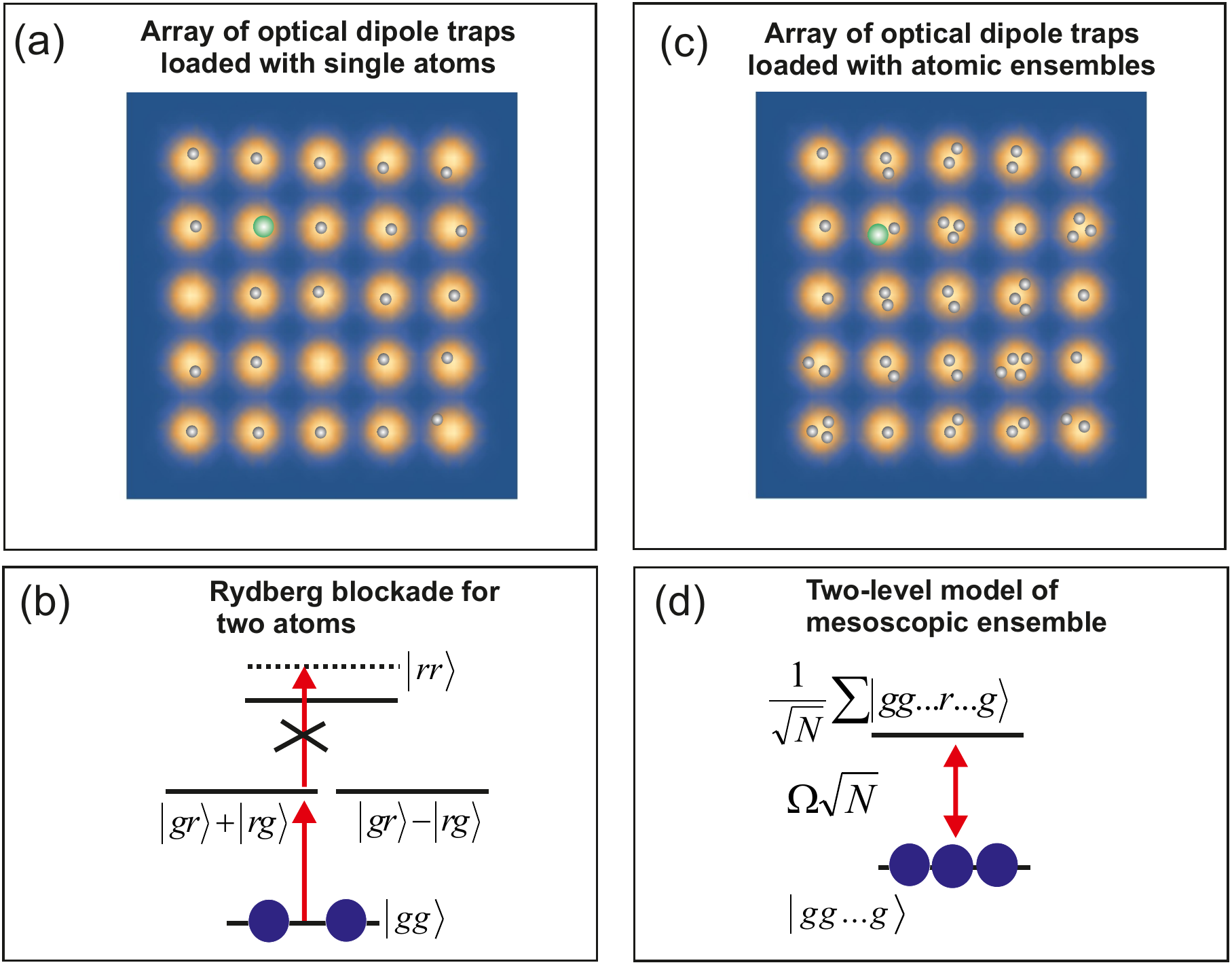}
\vspace{-.5cm}
\caption{
\label{Scheme}(Color online).
(a) Scheme of the quantum register based on individually addressed single atoms in the array of optical dipole traps. Laser pulses are used to excite atoms into the Rydberg state (shown as green circle). Simultaneous excitation of Rydberg atoms in the neighboring sites is  blocked; (b) Collective states of two interacting atoms. The shift of the collective energy level when both atoms are excited into the Rydberg state leads to suppression of double Rydberg excitation, known Rydberg blockade; (c)  Scheme of the quantum register based on individually addressed atomic ensembles in the array of optical dipole traps. Laser pulses are used to excite atoms into the Rydberg state. Only one atom in each site can be excited due to Rydberg blockade. Simultaneous excitation of Rydberg atoms in the neighboring sites is also blocked; (d) The mesoscopic ensemble of $N$ atoms in the Rydberg blockade regime can be considered as a two-level system with enhanced coupling to the laser field.
}
\end{figure}

Quantum computing is a challenging problem in modern physics~\cite{Nielsen2011}. Recently, great progress in quantum computing with superconductors~\cite{Benjamin2015,Richer2016,Rol2019} and ultracold ions~\cite{Ballance2016,Gaebler2016,Landsman2019} has been demonstrated, but their scalability to more than a hundred of qubits is questionable. Several other quantum systems (atoms, photons, quantum dots, etc.) remain promising alternatives for building a scalable quantum computer. Among them, ultracold neutral atoms meet all the DiVincenzo criteria for qubits~\cite{DiVincenzo2000,Saffman2010,Ryabtsev2016}. Hyperfine sublevels of the ground state of alkali-metal atoms with long lifetimes can be used as logical states of a qubit~\cite{Brennen1999, Jaksch2000}. The arrays of optical dipole traps [figure~\ref{Scheme}(a)] with single atom in each trap can be used as quantum registers of arbitrary dimensions~\cite{Xia2015,Barredo2018,Graham2019}. The qubits can be initialized via optical pumping to one of the hyperfine sublevels of the ground state. Single-qubit gates can be performed using microwave transitions between the hyperfine sublevels of the ground state or via two-photon laser Raman pulses~\cite{Saffman2005}. Two-qubit gates can be implemented using long-range interaction between Rydberg atoms~\cite{Jaksch2000,Lukin2001,Isenhower2010,Graham2019,Levine2019,Shi2017, Su2017,Petrosyan2017,Shi2018,Shi2018a, Shi2019,Li2018,Shi2019a}. The states of the atoms can be measured using resonance fluorescence~\cite{Saffman2010}. Despite the randomness in loading of each dipole trap, the atomic arrays can be reconfigured using movable optical tweezers, allowing building defect-free configurations~\cite{Barredo2018}. However, such quantum registers still suffer from single-atom loss, which is illustrated as empty slots in figure~\ref{Scheme}(a).

Recently, large quantum registers with ultracold neutral atoms have been experimentally demonstrated~\cite{Xia2015,Wang2016,Bernien2017,Graham2019}. However, building of high-fidelity quantum gates remains a challenging task. Two-qubit gates are the key element of a quantum computer. The most important are the controlled-NOT (CNOT) and controlled-phase (CZ) gates. Their successful implementation can be used in universal quantum computation. Two-qubit gates can be based on the effect of Rydberg blockade~\cite{Jaksch2000}, which   manifests itself as a suppression of the excitation of more than one atom in the ensemble by narrow-band laser radiation due to the shifts of the collective energy levels induced by long-range Rydberg-Rydberg interactions, as illustrated in figure~\ref{Scheme}(b) for two atoms. Rydberg blockade was successfully used in the experiment to build a CZ gate for ultracold neutral atoms with the fidelity above 0.97~\cite{Levine2019}. 

The mesoscopic atomic ensembles in the optical dipole traps [figure~\ref{Scheme}(c)] also represent qubits if these are controlled with Rydberg dipole blockade ensuring a single-atom laser excitation~\cite{Lukin2001,Zhao2018}. The advantage of these ensembles is the reduced sensitivity to atom losses in the trap~\cite{Saffman2010} and enhanced coupling to light~\cite{Lukin2001,Saffman2002}. Quantum information is encoded in collective states of each atomic ensemble.
An atomic ensemble in the regime of Rydberg blockade can be considered as a two-level system with the coupling to the laser radiation enhanced by a factor of \textit{N}, with \textit{N} being the number of atoms, as shown in figure~\ref{Scheme}(d). However,the fluctuations of the number of atoms in the ensemble due to random loading of the optical dipole traps substantially reduce the fidelity of quantum gates. 

We proposed several schemes of quantum gates which are based on adiabatic passage at Rydberg excitation to overcome the influence of the fluctuations of number of atoms on gate fidelity. Adiabatic excitation of a single Rydberg atom in the atomic ensemble in regime of Rydberg blockade can be used as an alternative technique of single-atom loading of large atomic arrays without the need for their spatial reconfiguration. We also considered adiabatic passage of Stark-tuned F\"{o}rster resonances for two Rydberg atoms. This allows building two-qubit controlled-phase gates based on weak long-range Rydberg interactions at large interatomic distances, in distinction with strong interactions and short distances required for dipole blockade.

Our approach is based on complex population and phase dynamics of adiabatic passage in multilevel atomic systems which can be of general interest for laser spectroscopy apart from applications for quantum information.
In the present work we review methods for coherently controlling Rydberg quantum states of atomic ensembles using Adiabatic Rapid Passage (ARP) and Stimulated Raman Adiabatic Passage (STIRAP). These methods are commonly used for population inversion in simple two-level and three-level systems. Several schemes of two-qubit gates based on adiabatic passage have been recently published~\cite{Wu2017}.

The paper is organized as follows. In Section~2 we review the theory of ARP and STIRAP in simple two-level and three-level systems. In Section~3 we discuss single-atom excitation in atomic ensembles in regime of Rydberg blockade and its possible applications for single-atom loading. In Section~4 we study phase dynamics during double adiabatic sequence for ARP and STIRAP. In Section~5 we discuss the schemes of quantum gates with mesoscopic atomic ensembles. Section~6 is devoted to the quantum gates based on adiabatic passage of the Stark-tuned F\"{o}rster resonances.


\section{Theory of adiabatic passage in atomic systems}
\subsection{Adiabatic Rapid Passage}

\begin{figure}[!t]
\begin{center}
\includegraphics[width=\columnwidth]{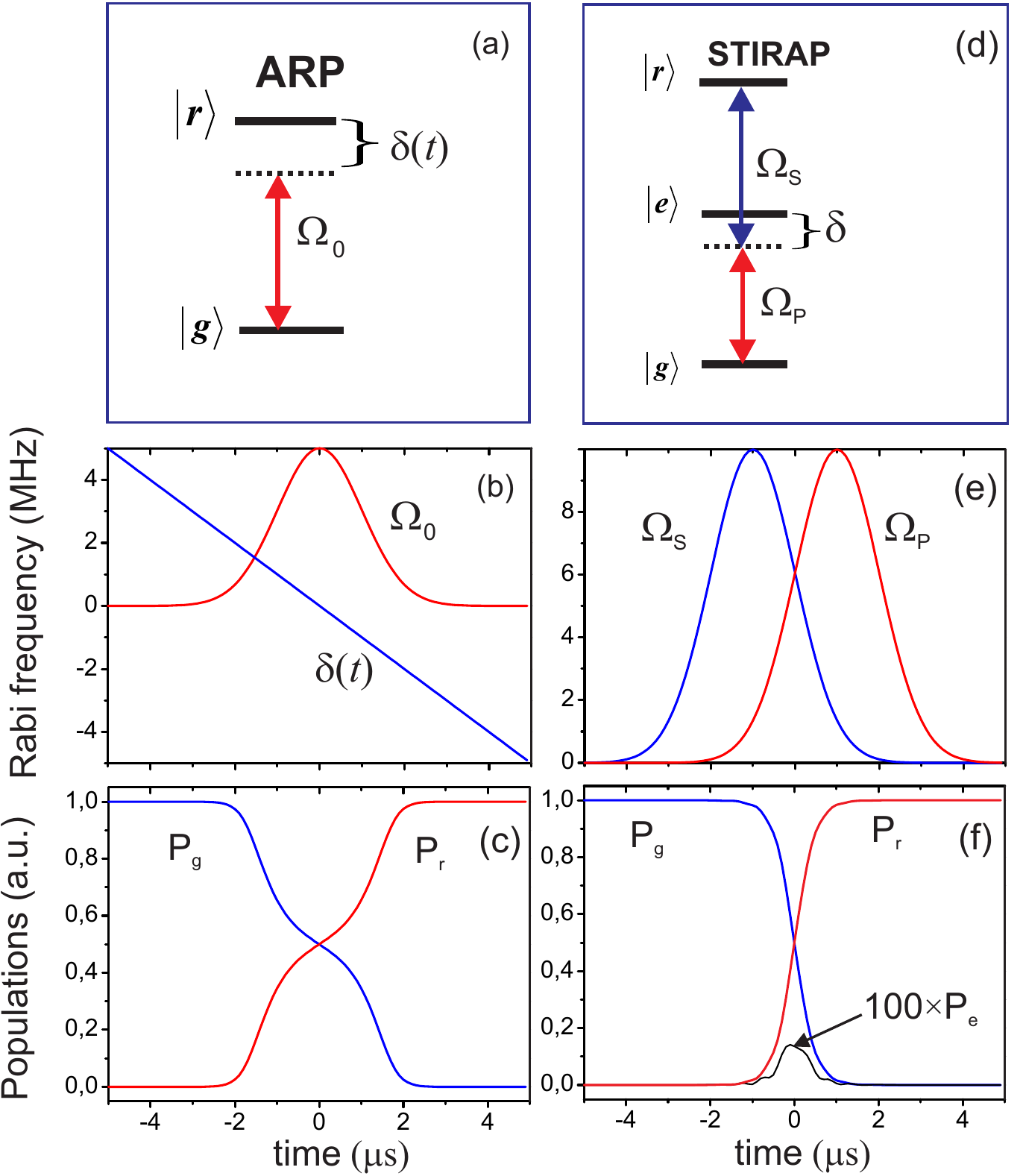}
\vspace{-.5cm}
\caption{
\label{ARP_STIRAP}(Color online).
(a) Two-level system interacting with laser radiation during ARP; (b) Time dependence of Rabi frequency $\Omega \left(t\right)$ and detuning $\delta \left(t\right)$ during ARP; (c) Time dependence of the population of the ground state $\ket{g}$ $\mathrm {P_g}$ and of the excited state $\ket{r}$ $\mathrm {P_r}$ during ARP; (d) Three-level system interacting with laser radiation during STIRAP; (e) Time-dependent Rabi frequencies $\Omega_S$ and $\Omega_P$ of the STIRAP pulses; (f) Time dependencies of the populations $\mathrm{P_g}$, $\mathrm{P_e}$ and $\mathrm{P_r}$ of the states $\ket{g}$, $\ket{e}$ and $\ket{r}$, respectively.
}
\end{center}
\end{figure}

Adiabatic rapid passage is commonly used for laser excitation of molecular levels because of the independence of transition probability of the Rabi frequency~\cite{Malinovsky2001,Malinovskaya2018,Kuznetsova2014}. Scheme of the two-level system interacting with laser radiation with time dependent Rabi frequency $\Omega\left(t\right)$ and detuning $\delta\left(t\right)$ is illustrated in figure~\ref{ARP_STIRAP}(a). The Hamiltonian for a two-level system with states $\ket{g}$ and $\ket{r}$, interacting with a chirped laser pulse (laser frequency and intensity change during the pulse), is written as

\be
\label{H_ARP}
\hat{\mathbf{H}}\left(t\right)=\frac{\hbar }{2} \left(\begin{array}{cc} {-\delta \left(t\right)} & {\Omega_0 \left(t\right)} \\ {\Omega_0 \left(t\right)} & {\delta \left(t\right)} \end{array}\right).
\ee

\noindent Here $\Omega_0 \left(t\right)$ is time-dependent Rabi frequency and $\delta \left(t\right)$ is time-dependent detuning from the resonance. In the field interaction representation the wavefunction is written as

\be
\label{Psi_ARP} 
\psi \left(t\right)=c_1\left(t\right)e^{i\omega t/2 }\ket{g} +c_{2} \left(t\right)e^{-i\omega t / 2}\ket{r}. 
\ee

\noindent Here $c_1\left(t\right)$ and $c_2\left(t\right)$ are probability amplitudes and $\omega$ is laser frequency. We define the 
time-dependent basis states to be $\ket{1\left(t\right)} =e^{i\omega t/2}\ket{g}$ and $\ket{2\left(t\right)} =e^{-i\omega t/2}\ket{r}$. In this basis the wavefunction is rewritten as follows:

\be
\label{Psi_ARP_RW} 
\ket{\psi \left(t\right)} =c_1 \left(t\right)\ket{ 1\left(t\right)} +c_2 \left(t\right)\ket{2\left(t\right)}.
\ee

\noindent To diagonalize the Hamiltonian, we rotate the basis:
\be
\label{ARP_Rotation_Basis}
\left(\begin{array}{c} {\ket{ I\left(t\right) } } \\ {\ket{ II\left(t\right) } } \end{array}\right)=\mathbf{T}\left(t\right)\left(\begin{array}{c} {\ket{ 1\left(t\right) } } \\ {\ket{ 2\left(t\right) } } \end{array}\right).    
\ee

\noindent Here $\ket{I\left(t\right)}$ and $\ket{II\left(t\right)}$ are semiclassical dressed states~\cite{Berman2011} and  $\mathbf{T}\left(t\right)$ is time-dependent unitary rotation matrix: 

\be
\label{ARP_Rotation}
\mathbf{T}\left(t\right)=\left(\begin{array}{cc} {\cos \theta \left(t\right)} & {-\sin \theta \left(t\right)} \\ {\sin \theta \left(t\right)} & {\cos \theta \left(t\right)} \end{array}\right).      
\ee 

\noindent where  $\theta \left(t\right)$ is a time-dependent mixing angle. The semiclassical dressed states are the superpositions:

\be
\label{DressedStates}
\begin{array}{l} {\ket{ I\left(t\right) } =\cos \theta \left(t\right)\ket{ 1\left(t\right) } -\sin \theta \left(t\right)\ket{ 2\left(t\right)} } \\ {\ket{II\left(t\right) } =\sin \theta \left(t\right)\ket{ 1\left(t\right) } +\cos \theta \left(t\right)\ket{ 2\left(t\right) } } \end{array}.     
\ee

\noindent To derive the equation for the probability amplitudes of dressed states  $\tilde{\mathbf{c}}$, we substitute the definition  $\tilde{\mathbf{c}}=\mathbf{T}\mathbf{c}$ into the Schr\"{o}dinger equation for the probability amplitudes $i\hbar \mathbf{\dot{c}=\hat{H}c}$. This results in

\be
\label{Transformation} 
i\hbar \dot{\tilde{\mathbf{c}}}=\mathbf{T}\hat{\mathbf{H}}\mathbf{T}^{+} \tilde{\mathbf{c}}-i\hbar \mathbf{T}\dot{\mathbf{T}}^{+} \tilde{\mathbf{c}}.
\ee

\noindent The matrix $\mathbf{T\hat{H}T^{+}}$ is diagonal if the mixing angle $\theta\left(t\right)$ obeys the following conditions:

\be
\label{MixingAngle}
\begin{array}{l} 
{tg\left[2\theta \left(t\right)\right]=\Omega_{0} \left(t\right) / \delta \left(t\right)} \\ 
{\sin \left[\theta \left(t\right)\right]=\sqrt{\frac{1}{2} \left(1-\frac{\delta \left(t\right)}{\Omega \left(t\right)} \right)} } \\
 {\cos \left[\theta \left(t\right)\right]=\sqrt{\frac{1}{2} \left(1+\frac{\delta \left(t\right)}{\Omega \left(t\right)} \right)} } \end{array}.     
\ee

\noindent Here $\Omega \left(t\right)=\sqrt{\Omega _{0}^{2} \left(t\right)+\delta \left(t\right)^{2} }$. This leads to:

\be
\label{H_diag}
\begin{array}{l} {\hat{\mathbf{H}}_{d} =\mathbf{T\hat{H}T^{+}} =\frac{\hbar }{2} \left(\begin{array}{cc} {-\Omega \left(t\right)} & {0} \\ {0} & {\Omega \left(t\right)} \end{array}\right)} \\ {\mathbf{T\dot{T}^{+}} =i\sigma_{y} \dot{\theta}} \end{array}.
\ee

\noindent  In the adiabatic approximation, when $\left|\dot{\Omega}_0\left(t\right)\right|/\Omega^2\left(t\right)\ll1$ and $\left|\dot{\delta}\left(t\right)\right|/\Omega^2\left(t\right)\ll1$ we can neglect the term proportional to $\dot{\theta}$. Then equation~(\ref{Transformation}) is rewritten as $i\hbar \dot{\tilde{\mathbf{c}}}=\hat{\mathbf{H}}_{d} \tilde{\mathbf{c}}$.  Its solution is 

\be
\label{ARP_Solution}
\begin{array}{l} {\tilde{c}_{1} \left(t\right)=\tilde{c}_{1} \left(0\right)\exp \left[-i\int\limits_{0}^{t}\Omega \left(t\right)dt \right]} \\ {\tilde{c}_{2} \left(t\right)=\tilde{c}_{2} \left(0\right)\exp \left[i\int\limits_{0}^{t}\Omega \left(t\right)dt \right]} \end{array}.    
\ee

\noindent 
The time dependence of Rabi frequency $\Omega_0 \left(t\right)$ and detuning $\delta \left(t\right)$ is shown in  figure~\ref{ARP_STIRAP}(b) for $\Omega_0 \left(t\right)=\Omega_0 \exp \left(-t^2 /2w^2  \right)$ with $\Omega_0/ \left(2\pi\right)=5$~MHz, $w=1\,\mu$s and $\delta\left(t\right)=\alpha t$ with $\alpha/(2\pi)=-1$~MHz/$\mu$s.   The system is initially in state $\ket{1\left(t\right)}$. For initial positive detuning $\delta \left(0\right)>0$ and $\Omega_0 \left(0\right)=0$ we find $\Omega \left(0\right)=\delta \left(0\right)$ and therefore $\theta \left(0\right)=0$. From Eq.~(\ref{MixingAngle}) the initial dressed state is  $\ket{I\left(t\right)}$ and $\tilde{c}_{1} \left(0\right)=1$. The time-dependent probability amplitudes are 

\be
\label{Amplitudes}
\begin{array}{l} {c_1 \left(t\right)=\tilde{c}_1 \left(t\right)\cos \theta \left(t\right)} \\ 
{c_2 \left(t\right)=-\tilde{c}_1 \left(t\right)\sin \theta \left(t\right)} \end{array}.
\ee

\noindent After the end of the adiabatic passage at time $T$ the detuning is negative $\delta \left(T\right)<0$ and $\Omega \left(T\right)=-\delta \left(T\right)$. Therefore the mixing angle $\theta \left(T\right)=\pi /2$, and the system ends in state $\ket{2\left(t\right)}$ which corresponds to the excited state $\ket{g}$.  

The time dependence of the population of the ground state $\ket{g}$ $\mathrm {P_g}$ and of the excited state $\ket{r}$ $\mathrm {P_r}$ are shown in figure~\ref{ARP_STIRAP}(c). ARP clearly results in population inversion in accordance with equation~(\ref{Amplitudes}).  

\subsection{Stimulated Raman Adiabatic Passage}

STIRAP is a two-photon adiabatic transition in a three-level system driven by a counterintuitive sequence of laser pulses~\cite{Bergmann1998}.
First, we consider a single atom with three energy levels denoted $\ket{g}$, $\ket{e}$ and $\ket{r}$ as shown in figure~\ref{ARP_STIRAP}(d), where $\ket{g}$ is a ground state, $\ket{e}$ is an intermediate excited state and $\ket{r}$ is a Rydberg state. The states $\ket{g}$ and $\ket{e}$ are coupled by a laser field with Rabi frequency $\Omega_{P} \left(t\right)$ and detuning $-\delta$ and the states $\ket{e}$ and $\ket{r}$ are coupled by a laser field with Rabi frequency $\Omega _{S} \left(t\right)$ and detuning $+\delta$ in such a way that both fields are tuned to the exact two-photon resonance for $\ket{g} \to \ket{r}$ transition. The Hamiltonian for a three-level system with states $\ket{g}$, $\ket{e}$ and $\ket{r}$ is written as~\cite{Bergmann1998,Berman2011}

\be
\label{H_STIRAP} 
\hat{\mathbf{H}}_{\mathrm{STIRAP}} \left(t\right)=\frac{\hbar }{2} \left(\begin{array}{ccc} {0} & {\Omega _{P} \left(t\right)} & {0} \\ {\Omega _{P} \left(t\right)} & {2\delta } & {\Omega _{S} \left(t\right)} \\ {0} & {\Omega _{S} \left(t\right)} & {0} \end{array}\right). 
\ee

\noindent The eigenvalues of the Hamiltonian are $\omega_D=0$ and $\omega_{A,B} =\pm \frac{\hbar}{2} \sqrt{\delta ^{2}+\Omega _{P}^{2} \left(t\right)+\Omega _{S}^{2} \left(t\right)} $. The eigenvalue $\omega _D=0$ corresponds to the dark state $\ket{D} =\cos \theta \left(t\right)\ket{g} -\sin \theta \left(t\right)\ket{r} $, where  $\theta \left(t\right)$ is a 
mixing angle and $\tan \theta \left(t\right)={\Omega _{P} \left(t\right) \mathord{\left/{\vphantom{\Omega _{P} \left(t\right) \Omega _{S} \left(t\right)}}\right.\kern-\nulldelimiterspace} \Omega _{S} \left(t\right)} $. We find $\cos \theta \left(t\right)={\Omega _{S} \left(t\right) \mathord{\left/{\vphantom{\Omega _{S} \left(t\right) \sqrt{\Omega _{S}^{2} \left(t\right)+\Omega _{P}^{2} \left(t\right)} }}\right.\kern-\nulldelimiterspace} \sqrt{\Omega _{S}^{2} \left(t\right)+\Omega _{P}^{2} \left(t\right)} } $ and $\sin \theta \left(t\right)={\Omega _{P} \left(t\right) \mathord{\left/{\vphantom{\Omega _{P} \left(t\right) \sqrt{\Omega _{S}^{2} \left(t\right)+\Omega _{P}^{2} \left(t\right)} }}\right.\kern-\nulldelimiterspace} \sqrt{\Omega _{S}^{2} \left(t\right)+\Omega _{P}^{2} \left(t\right)} } $. 

We consider the evolution of the probability amplitudes of the quantum states $\ket{g}$, $\ket{e}$ and $\ket{r}$ during the time interval $\left(-T,T\right)$ with $T=5\,\mu$s when the counterintuitive sequence of laser pulses is applied~\cite{Bergmann1998}. The time-dependent Rabi frequencies of the STIRAP pulses, shown in figure~\ref{ARP_STIRAP}(e), are expressed as

\bea
\label{STIRAP_Rabi}
\Omega _{S} \left(t\right)&=&\Omega _{S0} \exp \left[\left(t-t_{1} \right)^{2} /  2w^{2}  \right] \\ \nonumber
\Omega _{P} \left(t\right)&=&\Omega _{P0} \exp \left[\left(t-t_{2} \right)^{2} /  2w^{2}  \right] .
\eea

\noindent with $t_1 =-1\,\mu$s, $t_{2} =1\,\mu$s, $w=1\,\mu$s, $\Omega_{P0} /\left(2\pi\right)=\Omega_{S0} / \left(2\pi\right)=10$~MHz and $\delta /\left(2\pi\right)=10$~MHz.

The time dependencies of the populations $\mathrm{P_g}$, $\mathrm{P_e}$ and $\mathrm{P_r}$ of the states $\ket{g}$, $\ket{e}$ and $\ket{r}$, respectively, are shown in figure~\ref{ARP_STIRAP}(f).
Initially, $\cos \theta \left(t=-T\right)=1$, $\sin \theta \left(t=-T\right)=0$  and the state $\ket{g}$ is a dark state. During the adiabatic passage the atom remains in the dark state $\ket{D}$. After the end of the pulse sequence $\cos \theta \left(t=-T\right)=0$ and $\sin \theta \left(t=-T\right)=1$, and the atom is transferred to the state $\ket{r}$.


\section{Deterministic single-atom excitation}

In the regime of Rydberg blockade only one atom in the mesoscopic ensemble can be excited into Rydberg state due to the shift of the collective energy levels, induced by Rydberg interactions. Only symmetric collective states can be excited. Therefore, such an ensemble can be considered as a two-level system with two collective states $\ket{\bar 0}$ and $\ket{\bar r} $ :

\bea
\ket{\bar 0} = \ket{000...000}, \\\nonumber
\ket{\bar r} =\frac{1}{\sqrt{N}}\sum_{j=1}^N \ket{000 ... r_j ...000}. 
\eea

\noindent The second state is a symmetric superposition of all possible states when $j$th atom is excited into Rydberg state. The coupling for these collective states is $\sqrt{N}$ enhanced, compared to a single-atom case. Therefore, when the number $N$ of trapped atoms is random and unknown, coherent population inversion by a single $\pi$ laser pulse becomes impossible. We have proposed to use ARP or STIRAP to deterministically excite a single Rydberg atom in the mesoscopic atomic ensemble in the regime of Rydberg blockade~\cite{Beterov2011}. 

\begin{figure}[!t]
\includegraphics[width=\columnwidth]{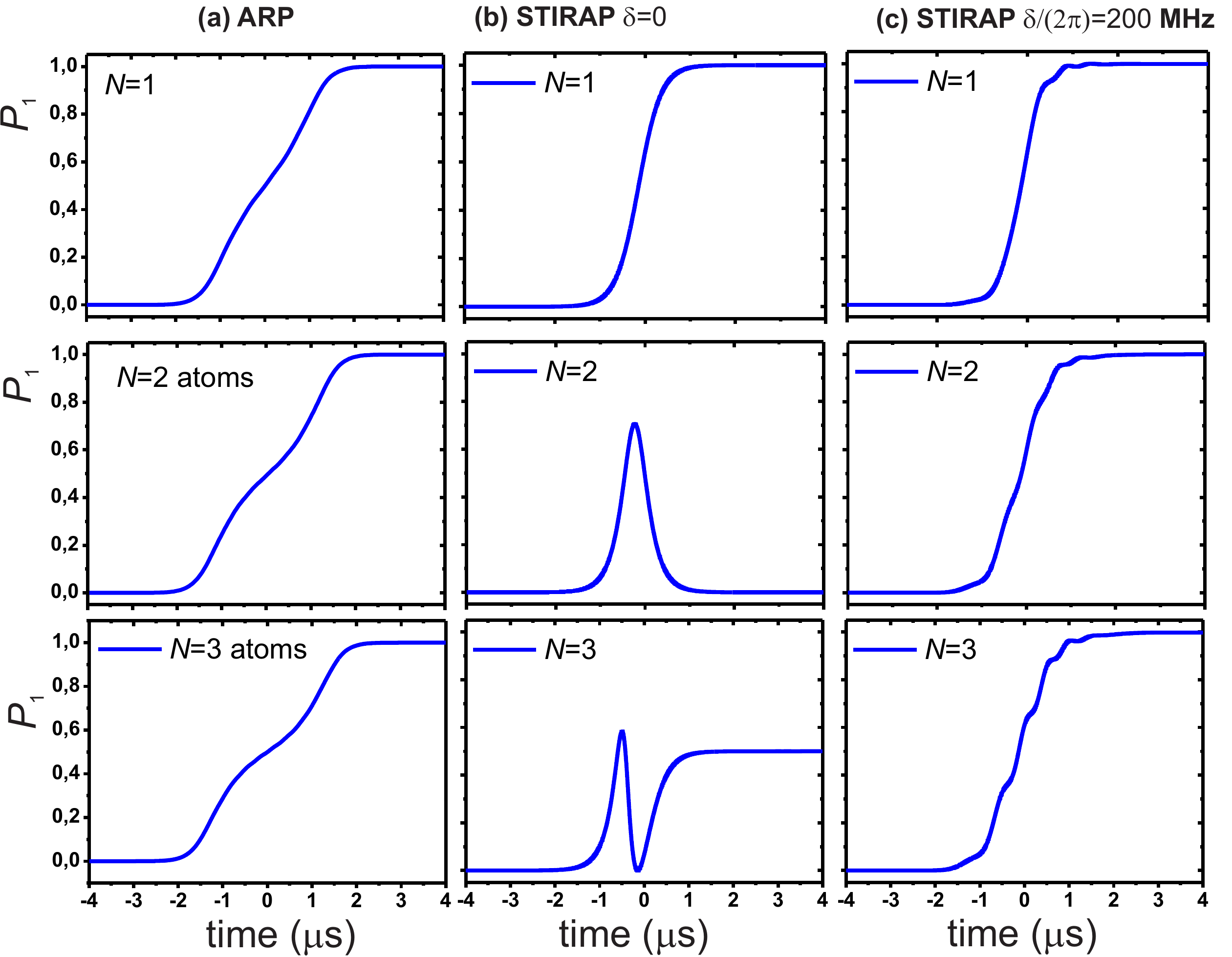}
\vspace{-.5cm}
\caption{
\label{Deterministic}(Color online).
Calculated time dependence of the probability of single-atom Rydberg excitation for $N=1-3$ atoms (top to bottom).
(a) ARP with the chirp rate is  $\alpha/\left(2\pi\right)=1$~MHz/$\mu$s and Rabi frequency is $\Omega_1/\left(2 \pi \right)=2$~MHz;
(b) STIRAP with $\Omega_{P0}/(2\pi) = 30~\rm MHz,$ $\Omega_{S0}/(2\pi) = 40~\rm MHz$, $\delta/(2\pi) = 0$;
(c) STIRAP with $\Omega_{P0}/(2\pi) = 30~\rm MHz,$ $\Omega_{S0}/(2\pi) = 40~\rm MHz$, $\delta/(2\pi) = 200$~MHz.
}
\end{figure}

We have numerically calculated the probability of single-atom Rydberg excitation in the mesoscopic ensembles with $N<10$ atoms in the regime of Rydberg blockade for a linearly-chirped Gaussian laser pulse and STIRAP sequence.  Calculations were performed using the Schr\"odinger equation, neglecting spontaneous emission, and assuming perfect blockade so only states with at most a single Rydberg excitation were included. We considered ARP by a single laser pulse with  $\Omega_0 \left(t\right)=\Omega_0 \exp \left(-t^2 /2w^2  \right)$ for $\Omega_0/ \left(2\pi\right)=2$~MHz, $w=1\,\mu$s and $\delta\left(t\right)=\alpha t$ with $\alpha/(2\pi)=-1$~MHz/$\mu$s. 
The time dependence of the probability $\mathrm{P_r}$ of single-atom Rydberg excitation for ARP is illustrated in panel~(a) of figure~\ref{Deterministic} for $N$=1,2 and 3 atoms. The probability of single-atom excitation for ARP was found to be independent of the number of atoms.

We have found that for STIRAP the population dynamics is more complex. 
The STIRAP sequence used Gaussian pulses described by equation~(\ref{STIRAP_Rabi}) with $t_1 =-1\,\mu$s, $t_{2} =1\,\mu$s, $w=1\,\mu$s,  $\Omega_{P0} /\left(2\pi\right)=40$~MHz and $\Omega_{S0} /\left(2\pi\right)=30$~MHz. If $\delta /\left(2\pi\right)=0$, STIRAP results in population inversion in a single-atom system, as the system remains in the dark state, described in Section~2.2. However, for two atoms in the regime of Rydberg blockade inital collective state $\ket{\bar 0}$ is not a dark state anymore. The population transfer between ground and excited collective state becomes impossible, as clearly seen in panel~(b) of figure~\ref{Deterministic}. 

The two-photon STIRAP becomes identical to ARP, if the detuning from the intermediate excited state is substantially large to eliminate the intermediate state from the equations. The time-dependent light shifts in a three-level system play the role of the time-dependent detuning in a two-level system. The numerically calculated  time dependencies of the population of collective state   $\ket{\bar r} $ during STIRAP with $\delta /\left(2\pi\right)=200$~MHz are shown in panel~(c) of figure~\ref{Deterministic}. The probability of single-atom Rydberg excitation is independent of the number of atoms in the ensemble.

The properties of STIRAP for blockaded ensemble can be partly explained by analysis of the simplest two-atom example~\cite{Beterov2017}.
The Hamiltonian for two three-level atoms in the regime of Rydberg blockade is written for eight collective states $\ket{gg}$, $\ket{ge}$, $\ket{gr}$, $\ket{eg}$, $\ket{ee}$, $\ket{er}$, $\ket{rg}$, $\ket{re}$ of a quasimolecule, consisting of two interacting atoms.  We take into account the effect of Rydberg blockade by removing the collective state $\ket{rr}$ with double Rydberg excitation from the Hamiltonian~\cite{Moller2008}:

\begin{widetext}
\be
\label{H_2STIRAP}
\hat{\mathbf{H}}_{\mathrm{2STIRAP}} \left(t\right)=\frac{\hbar }{2} \left(\begin{array}{cccccccc} {0} & {\Omega _{P} \left(t\right)} & {0} & {\Omega _{P} \left(t\right)} & {0} & {0} & {0} & {0} \\ {\Omega _{P} \left(t\right)} & {2\delta \left(t\right)} & {\Omega _{S} \left(t\right)} & {0} & {\Omega _{P} \left(t\right)} & {0} & {0} & {0} \\ {0} & {\Omega _{S} \left(t\right)} & {0} & {0} & {0} & {\Omega _{P} \left(t\right)} & {0} & {0} \\ {\Omega _{P} \left(t\right)} & {0} & {0} & {2\delta \left(t\right)} & {\Omega _{P} \left(t\right)} & {0} & {\Omega _{S} \left(t\right)} & {0} \\ {0} & {\Omega _{P} \left(t\right)} & {0} & {\Omega _{P} \left(t\right)} & {4\delta \left(t\right)} & {\Omega _{S} \left(t\right)} & {0} & {\Omega _{S} \left(t\right)} \\ {0} & {0} & {\Omega _{P} \left(t\right)} & {0} & {\Omega _{S} \left(t\right)} & {2\delta \left(t\right)} & {0} & {0} \\ {0} & {0} & {0} & {\Omega _{S} \left(t\right)} & {0} & {0} & {0} & {\Omega _{P} \left(t\right)} \\ {0} & {0} & {0} & {0} & {\Omega _{S} \left(t\right)} & {0} & {\Omega _{P} \left(t\right)} & {2\delta \left(t\right)} \end{array}\right).
\ee
\end{widetext}

\begin{center}
\begin{figure}[!t]
 \center
\includegraphics[width=\columnwidth]{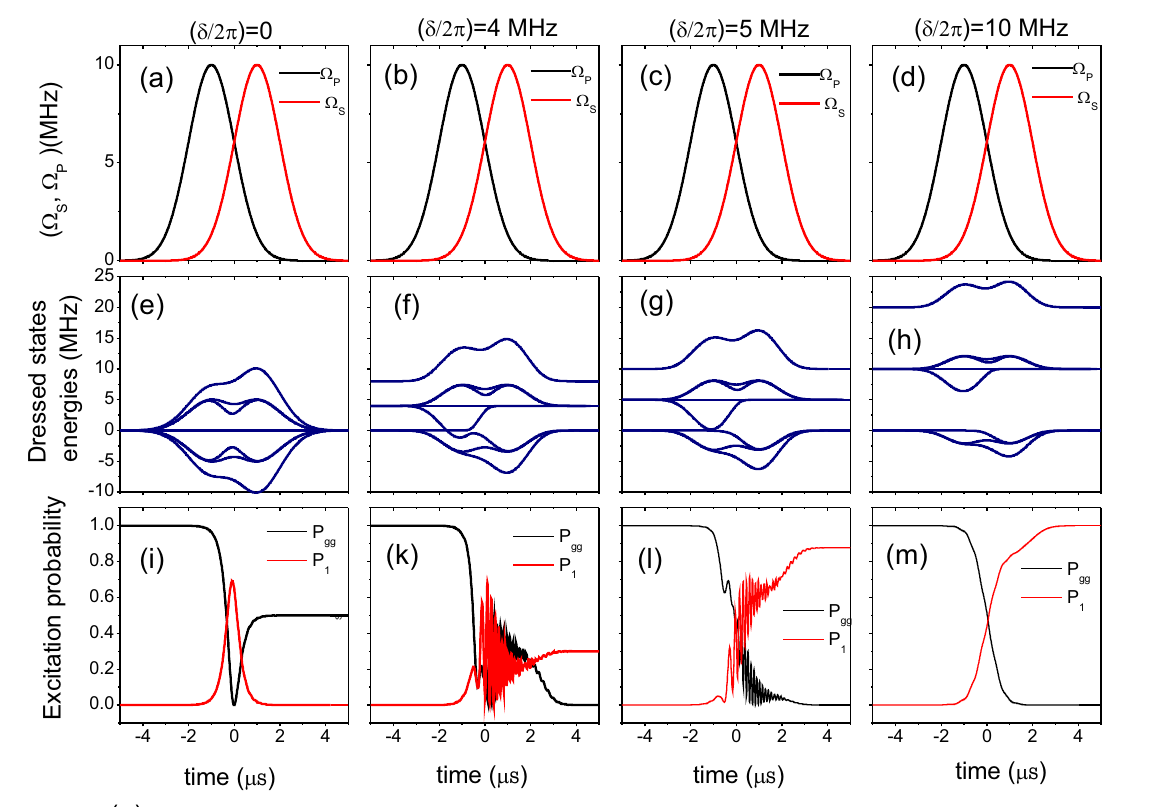}
\vspace{-.5cm}
\caption{
\label{Regimes}(Color online).
(a)-(d) Time sequence of STIRAP pulses; (e)-(h) Eigenvalues of the two-atom Hamiltonian for different detunings from the intermediate state $\delta  / 2\pi =0 $, $\delta  / 2\pi =4$~MHz, $\delta  / 2\pi =5 $MHz, and $\delta  / 2\pi =10$~MHz, respectively; (i)-(m) Time dependencies of the probability $P_{gg}$, and the probability to excite a single Rydberg atom $P_1$ for different detunings from the intermediate state $\delta  / 2\pi =0 $, $\delta  / 2\pi =4$~MHz, $\delta  / 2\pi =5$~MHz, and $\delta  /2\pi =10$~MHz, respectively.
}
\end{figure}
\end{center}

\noindent 
We calculated the eigenvalues for the Hamiltonian with the Rabi frequencies of equation~(\ref{STIRAP_Rabi}) [figures~\ref{Regimes}(a)-(d)] for four different constant detunings from the intermediate state ${\delta  \mathord{\left/{\vphantom{\delta  \left(2\pi \right)}}\right.\kern-\nulldelimiterspace} \left(2\pi \right)} =0$,   ${\delta  \mathord{\left/{\vphantom{\delta  \left(2\pi \right)}}\right.\kern-\nulldelimiterspace} \left(2\pi \right)} =4$~MHz, ${\delta  \mathord{\left/{\vphantom{\delta  \left(2\pi \right)}}\right.\kern-\nulldelimiterspace} \left(2\pi \right)} =5$~MHz, ${\delta  \mathord{\left/{\vphantom{\delta  \left(2\pi \right)}}\right.\kern-\nulldelimiterspace} \left(2\pi \right)} =10$~MHz, shown in figures~\ref{Regimes}(e)-(h), respectively. The calculated time dependences of the probability $P_{gg}$ of the collective ground state $\ket{gg}$ and of the probability $P_1$ of the collective state with single Rydberg excitation $\frac{1}{\sqrt{2}}\left(\ket{gr}+\ket{rg}\right)$ are shown in figures~\ref{Regimes}(i)-(m). At zero detuning from the intermediate state there is a dark state with zero eigenvalue, as shown in figure~\ref{Regimes}(e)~\cite{Moller2008}. STIRAP does not provide a Rydberg excitation  after the end of the pulse sequence in this particular case~\cite{Moller2008}. On contrary, for non-zero detuning from the intermediate state there is no dark state with zero eigenvalue, as shown in figures~\ref{Regimes}(f)-(h).  The switching between the regimes of single-atom Rydberg excitation occurs around ${\delta  \mathord{\left/{\vphantom{\delta  \left(2\pi \right)}}\right.\kern-\nulldelimiterspace} \left(2\pi \right)} =5$~MHz, as shown in figures~\ref{Regimes}(k) and \ref{Regimes}(l). For ${\delta  \mathord{\left/{\vphantom{\delta  \left(2\pi \right)}}\right.\kern-\nulldelimiterspace} \left(2\pi \right)} =10$~MHz STIRAP results in the deterministic single-atom Rydberg excitation (independent of \textit{N}) after the end of the adiabatic passage, as shown in figure~\ref{Regimes}(m).

\begin{figure}[!t]
\includegraphics[width=\columnwidth]{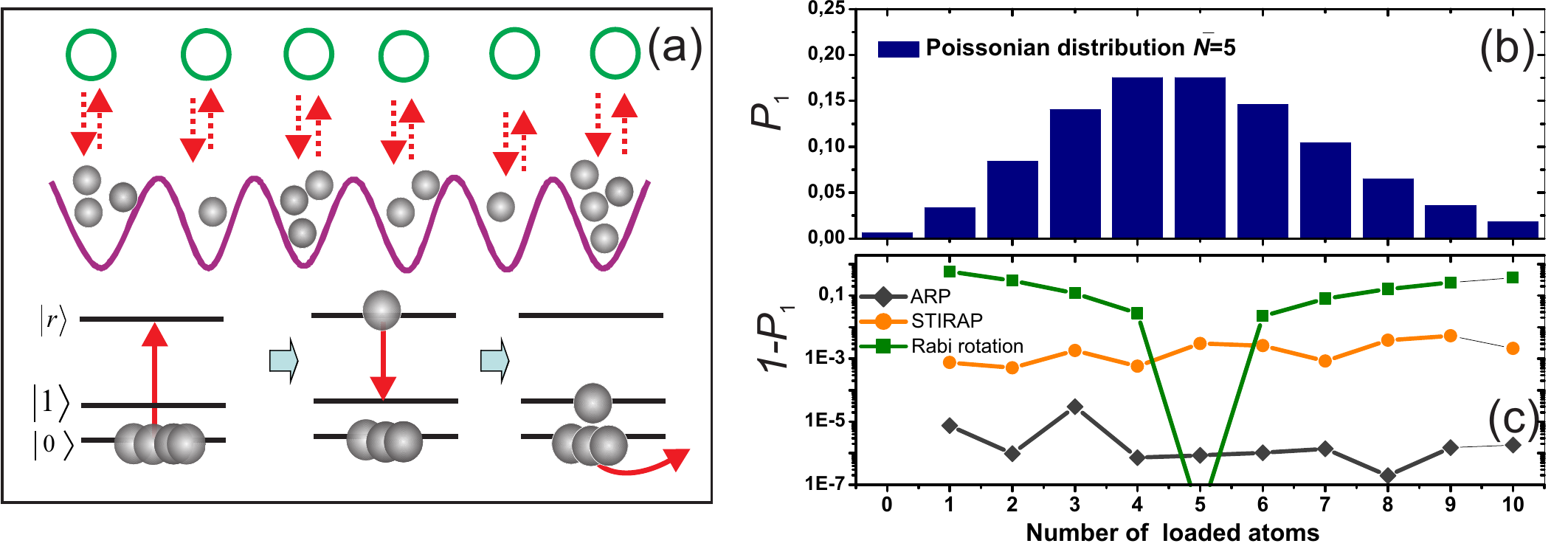}
\vspace{-.5cm}
\caption{
\label{Loading}(Color online).
(a) The scheme of single-atom loading using deterministic single-atom Rydberg excitation;
(b) The Poissonian statistics of loading of optical dipole trap witn the average number of atoms $\bar N=5$;
(c) The fidelity of single-atom excitation for a single-photon $\pi$ rotation with the area optimized for $N=5$ atoms compared
to STIRAP or ARP pulses.
}
\end{figure}

This technique of single-atom excitation can be used for deterministic single-atom loading, proposed in~\cite{Saffman2002}, when one of the atoms is deterministically transfered between the hyperfine sublevels of the ground state through temporarily Rydberg excitation in the blockade regime, while all atoms remained at the inititally populated hyperfine sublevel are removed from the optical dipole trap by an additional laser pulse, as shown in figure~\ref{Loading}(a). A similar problem has been recently addressed in~\cite{Petrosyan2013}.

The probability of loading $N$ noninteracting atoms in a small  optical or magnetic trap is described, in general, by
Poissonian statistics. For $\bar{N}=5$ the probability to load zero atoms is 0.0067, as shown in figure~\ref{Loading}(b), which is  small enough to create a large quantum register with a small number of defects. Figure~\ref{Loading}(c) shows a comparison of the
fidelity of single-atom excitation for a single-photon $\pi$ rotation with the area optimized for $N=5$ atoms compared
to STIRAP or ARP pulses. We see that the adiabatic pulses reduce the population error by up to several orders of
magnitude for a wide range of $N$.

The fidelity of STIRAP with Gaussian pulses is worse than the fidelity of ARP. Commonly used STIRAP techniques with Gaussian pulses usually provide the infidelity larger than 10\textsuperscript{-4} even in theory. The fidelity of the population transfer can be improved by optimization of the shapes of STIRAP pulses, as proposed in Ref.~\cite{Vasiliev2009}.

\bea
\label{VitanovRabi}
\Omega _{P} \left(t\right)&=&\Omega_V F\left(t-t_0\right)\mathrm{cos}\left[\frac{\pi} 2f\left(t-t_0\right)\right] \\ \nonumber
\Omega _{S} \left(t\right)&=&\Omega_V F\left(t-t_0\right)\mathrm{sin}\left[\frac{\pi } 2f\left(t-t_0\right)\right].
\eea

\begin{center}
\begin{figure}[!t]
 \center
\includegraphics[width=\columnwidth]{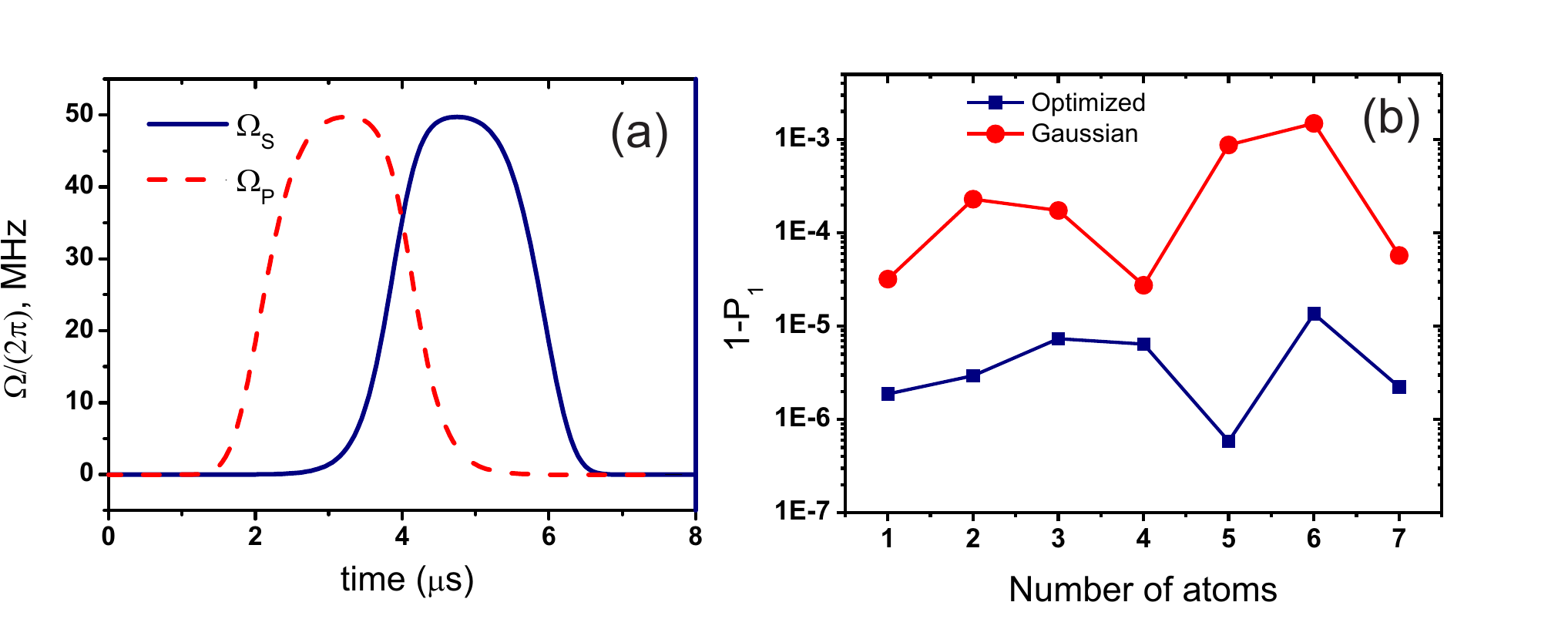}
\vspace{-.5cm}
\caption{
\label{Vitanov}(Color online).
(a) Shapes of the optimized STIRAP pulses~\cite{Vasiliev2009}; (b) Comparison of the error $1-P_1$ of population transfer for Gaussian and optimized STIRAP sequence.
}
\end{figure}
\end{center}

\noindent Here  $F\left(t\right)=\exp\left[-\left(t/T_0\right)^{2n}\right]$ and  $f\left(t\right)=\left[1+\exp\left(-\mathit{\lambda t}/\tau\right)\right]^{-1}$ . Following Ref.~\cite{Vasiliev2009}, we have chosen  $T_0=2\tau$,  $n=3$, and  $\lambda =4$. In our calculations the Rabi frequency for both pulses is $\Omega _V/\left(2\pi \right)=50$ MHz, detuning from the intermediate state is $\delta /\left(2\pi \right)=200$ MHz, and $T_0=2\;\mathit{\mu s}$ is the time parameter for a hypergaussian function $F\left(t\right)$ which determines the pulse duration. The positions of the pulses are defined by  $t_0=4\,\mu s$. The shapes of the STIRAP pulses are shown in figure~\ref{Vitanov}(a).


We have compared the fidelity of population inversion of the optimized STIRAP scheme with the conventional Gaussian pulses from equation~(\ref{STIRAP_Rabi}) with $\Omega _{P0}=\Omega _{S0}=\Omega _V$, $t_1=-1\;\mathit{\mu s}$,  $t_2=1\;\mathit{\mu s}$ and  $w =1\;\mathit{\mu s}$.

Comparison of the numerically calculated fidelity of single-atom Rydberg excitation in the atomic ensemble consisting of \textit{N} atoms for Gaussian and optimized pulses is shown in figure~\ref{Vitanov}(b). We have solved a Schr\"odinger equation for the probability amplitudes in a quasimolecule which consists of \textit{N} three-level atoms, interacting with two laser fields. The perfect Rydberg blockade was considered in the simulations by removing all quasimolecular states with more than one Rydberg excitation. The finite lifetimes of intermediate and Rydberg states have not been taken into account (this assumes short interaction times compared to lifetimes). The optimized pulse shapes allow substantial reduction of the infidelity of single-atom Rydberg excitation, which is kept below 10\textsuperscript{-5} for almost all cases, as shown in figure~\ref{Vitanov}(b).

\section{Phase accumulation during double adiabatic sequences}

\subsection{Adiabatic Rapid Passage}

\begin{center}
\begin{figure}[!t]
 \center
\includegraphics[width=\columnwidth]{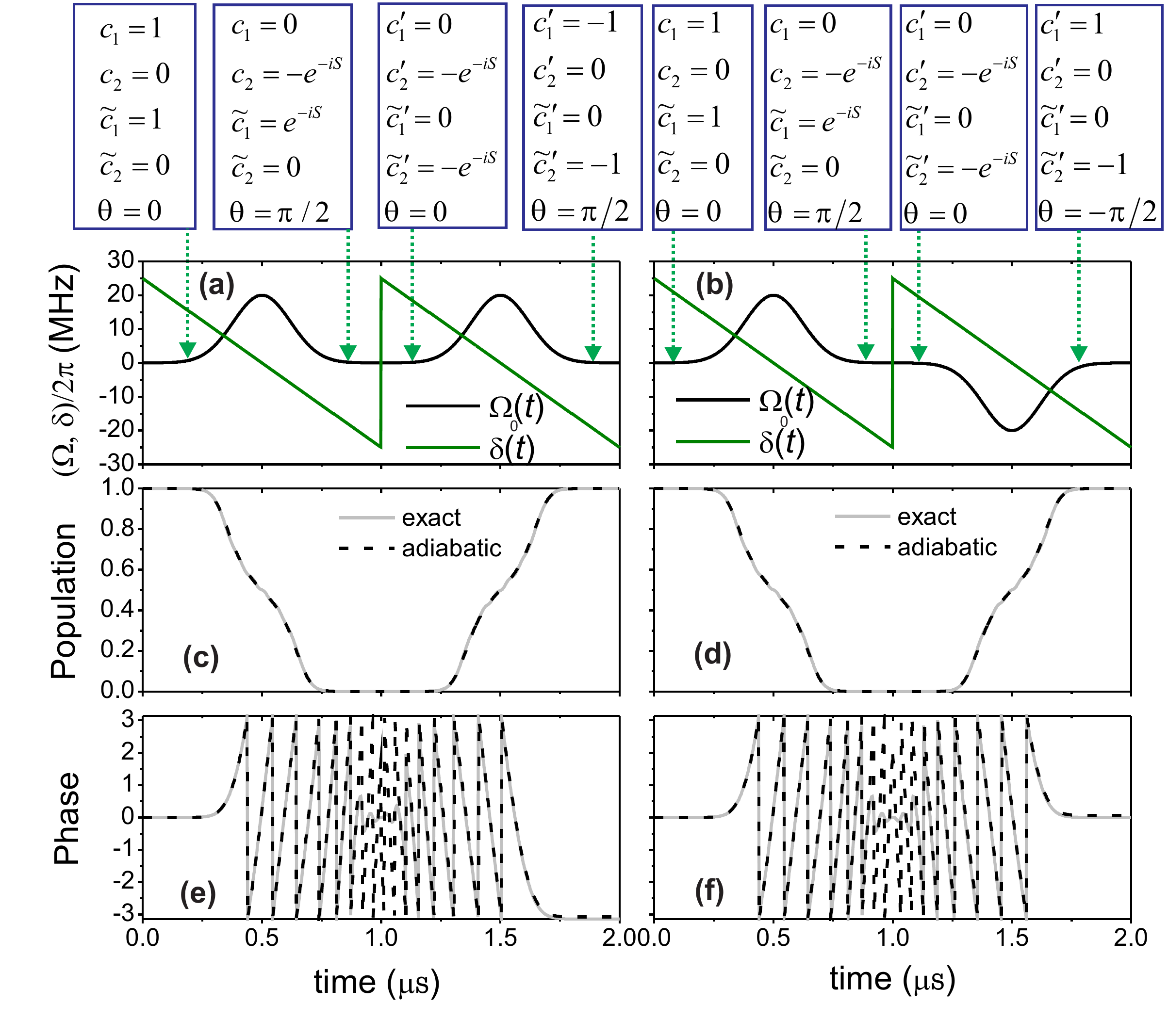}
\vspace{-.5cm}
\caption{
\label{Analyt}(Color online).
 Scheme of deterministic phase accumulation during a double adiabatic passage in a two-level quantum system. The phase shift is $\pi$ for the left-hand panel and zero for the right-hand panel. The dynamics of probability amplitudes $c_{1} \left(t\right)$,$c_{2} \left(t\right)$, $c'_{1} \left(t\right)$, $c'_{2} \left(t\right)$ of states  $\ket{1\left(t\right),2\left(t\right)} $  and of probability amplitudes $\tilde{c}_{1} \left(t\right)$, $\tilde{c}_{2} \left(t\right)$, $\tilde{c'}_{1} \left(t\right)$, $\tilde{c'}_{2} \left(t\right)$ of semiclassical dresses states $\ket{ I\left(t\right),II\left(t\right)}$ is shown schematically.  (a), (b) Time dependences of Rabi frequency $\Omega \left(t\right)$ and of detuning $\delta \left(t\right)$; (c), (d) Numerically calculated time dependences of the population of  initial state $\ket{1\left(t\right)}$ compared with the calculations in the adiabatic approximation. (e), (f) Numerically calculated time dependencies of the phase of  initial state $\ket{1\left(t\right)}$ compared with calculations in the adiabatic approximation. 
}
\end{figure}
\end{center}

Population transfer by adiabatic passage in two-level and three-level systems has been extensively studied for years. However, the phase accumulation during adiabatic passage was much less discussed. In our previous works we have shown that the phases of the collective atomic states can be efficiently controlled by double adiabatic sequences which return the atomic system to the initial state.
The simplest example of phase accumulation after double adiabatic passage is a sequence of two linearly chirped Gaussian pulses, illustrated in figure~\ref{Analyt}. The double adiabatic sequence  starts at \textit{t}=0. The time dependence of Rabi frequency $\Omega_0\left(t\right)$ and detuning $\delta\left(t\right)$ is illustrated in Fig.~\ref{Analyt}(a). The system is initially in state $\ket{1\left(t\right)}$. For initial positive detuning $\delta \left(0\right)>0$ and $\Omega_0 \left(0\right)=0$ we find $\Omega \left(0\right)=\delta \left(0\right)$ and therefore the initial mixing angle $\theta \left(0\right)=0$. From equation~(\ref{DressedStates}) the initial dressed state is  $\ket{I\left(t\right)}$ and $\tilde{c}_{1} \left(0\right)=1$. The time-dependent probability amplitudes are described by the equation~(\ref{Amplitudes}).

After the end of the first adiabatic passage at time $T$ the detuning is negative $\delta \left(T\right)<0$ and $\Omega \left(T\right)=-\delta \left(T\right)$. Therefore the mixing angle $\theta \left(T\right)=\pi /2$, and the system ends in state $\ket{2\left(t\right)}$ with $c_2 \left(T\right)=-\tilde{c}_1 \left(T\right)=-\exp\left[-i\int_0^{T}\Omega\left(t\right)dt \right]=-\exp \left[-i S\right]$. Here $S$ is a generalized pulse area.  

We denote the mixing angle and the probability amplitudes for the second adiabatic passage as $\theta'$, $c_1'\left(t\right)$, $c_2'\left(t\right)$, $\tilde{c}_1'\left(t\right)$, $\tilde{c}_2'\left(t\right)$. At the beginning of the second adiabatic passage the detuning is positive $\delta \left(T\right)>0$ and $\theta'\left(T\right)=0$. At time $t=T$ the system is in state $\ket{2\left(t\right)}$. From equation~(\ref{DressedStates}) the dressed state is now $\ket{II\left(t\right)}$. The probability amplitude $c_2\left(t\right)$ of state $\ket{2\left(t\right)}$ is constant around $t=T$ due to the absence of interaction with the laser field. Therefore, the initial probability amplitude of dressed state  $\ket{II\left(t\right)}$ is $\tilde{c}'_2 \left(T\right)=c_2\left(T\right)=-\tilde{c}_{1} \left(T\right)$. During the second adiabatic passage the time-dependent probability amplitudes are expressed similarly to equation~(\ref{Amplitudes}):

\be
\label{AmplitudesPrime}
\begin{array}{l} {c'_1 \left(t\right)=\tilde{c}'_2 \left(t\right)\sin \theta '\left(t\right)} \\
 {c'_2 \left(t\right)=\tilde{c}'_2 \left(t\right)\cos \theta '\left(t\right)} \end{array}.
\ee

\noindent From equation~(\ref{ARP_Solution}) the probability amplitude of dressed state $\ket{II\left(t\right)}$ is $\tilde{c}'_2 \left(t\right)=\tilde{c}'_2 \left(T\right)\exp \left[-i\int _{T}^{t}\Omega \left(t\right)dt \right]$. After the end of the second adiabatic passage the mixing angle is $\theta '\left(2T\right)=\pi /2$ and the system ends in state $\ket{1\left(t\right)}$ with probability amplitude 

\bea
\label{AmplitudesFinal} c'_{1}\left(2T\right)&=&\tilde{c}'_2 \left(2T\right)=\\
&=&-\exp \left[i\int\limits_{T}^{2T}\Omega \left(t\right)dt \right]\exp \left[-i\int\limits_{0}^{T}\Omega \left(t\right)dt \right]. \nonumber
\eea

\noindent For two identical laser pulses we find $c'_{1} \left(2T\right)=-1$. This corresponds to a $\pi$ phase shift which can be used for implementation of a CZ gate~\cite{Saffman2019}

This $\pi$ phase shift can be compensated if the second laser pulse has the opposite sign of Rabi frequency $\Omega _{0} \to -\Omega _{0}$ (which is equivalent to a $\pi$ phase shift of the laser pulse), as shown in Fig.~\ref{Analyt}(b). To diagonalize the Hamiltonian for the second adiabatic passage, we modify the equation~(\ref{MixingAngle}):

\be
\label{MixingAngleSecond}
\begin{array}{l} 
{tg\left[2\theta \left(t\right)\right]=-\Omega_{0} \left(t\right) / \delta \left(t\right)} \\ 
{\sin \left[\theta \left(t\right)\right]=-\sqrt{\frac{1}{2} \left(1-\frac{\delta \left(t\right)}{\Omega \left(t\right)} \right)} } \\
 {\cos \left[\theta \left(t\right)\right]=\sqrt{\frac{1}{2} \left(1+\frac{\delta \left(t\right)}{\Omega \left(t\right)} \right)} } \end{array}.     
\ee

\noindent In this case after the end of the second adiabatic passage $\theta'\left(2T\right)=-\pi /2$ and $c'_1\left(2T\right)=1$. 

\subsection{STIRAP}

Now we consider the dynamical phase accumulation during STIRAP in an ensemble of two strongly interacting atoms. 

Due to the absence of the dark state, the dynamical phase is accumulated during adiabatic passage. This is undesirable for quantum information with mesoscopic atomic ensembles due to the dependence of the phase on the number of interacting atoms~\cite{Beterov2013}. In our previous works~\cite{Beterov2011,Beterov2013,Beterov2014,Beterov2016} we have found that a double STIRAP sequence with the switched sign of the detuning from the intermediate state between two STIRAP sequences can be used to suppress the dynamical phase accumulation.

\begin{center}
\begin{figure}[!t]
 \center
\includegraphics[width=\columnwidth]{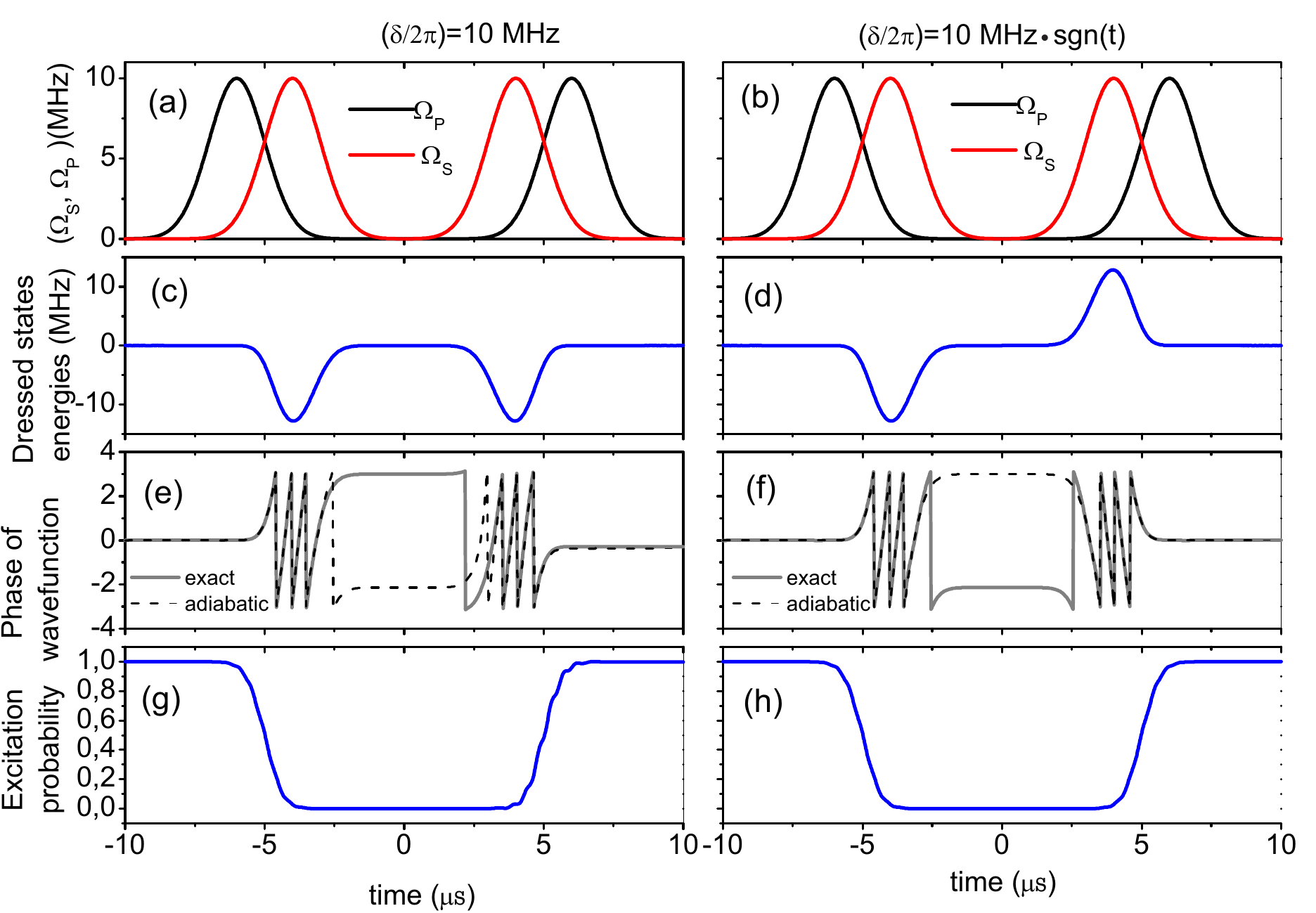}
\vspace{-.5cm}
\caption{
\label{STIRAPreverse}(Color online).
(left-hand panel) Double STIRAP  with detuning from the intermediate state $\delta  / 2\pi =10$~MHz; (right-hand panel)  Double STIRAP  with detuning from the intermediate state $\delta  / 2\pi =10 \,{\rm MHz}\times {\rm sgn}\left({\rm t}\right)$; (a), (b) Time sequence of double STIRAP pulses; (c), (d) Time dependencies of the eigenvalue of the two-atom Hamiltonian which corresponds to the eigenstate when the two-atom system is initially in state $\ket{gg}$; (e),(f) Comparison of the numerically calculated time dependence of the phase of the state $\ket{gg}$ (solid line) with the adiabatic approximation (dashed line); (g),(h) Time dependencies of the population of the ground state $\ket{gg}$.
}
\end{figure}
\end{center}

To explain this, we consider a double STIRAP sequence during  the time interval $\left(-T,T\right)$ with $T=10\,\mu$s. The shapes of laser pulses, shown in figures~\ref{STIRAPreverse}(a), (b), are described as
\bea
\label{STIRAP_Double}
\Omega _{S} \left(t\right)&=&\Omega _{S0} \exp \left[\frac{\left(t-t_{1} \right)^2}{  2w^2}  \right]+\Omega _{S0}  \exp \left[\frac{\left(t+t_{1} \right)^2}{  2w^2}  \right]  \nonumber \\
\Omega _{P} \left(t\right)&=&\Omega _{P0} \exp \left[\frac{\left(t-t_{2} \right)^2} {  2w^2 } \right]+\Omega _{P0}  \exp \left[\frac{\left(t+t_{2} \right)^2}{  2w^2}  \right] \nonumber \\
\eea

\noindent
with $\Omega_{P0}/(2\pi) =\Omega_{S0}/(2\pi)=10$~MHz, $t_1=4\,\mu$s, $t_2=6\,\mu$s,  and constant detuning  $\delta/(2\pi) =10$~MHz (left-hand panel of figure~\ref{STIRAPreverse}) and   $\delta /(2\pi)=10 \times \mathrm{sgn}\left(t\right)$~MHz (right-hand panel of figure~\ref{STIRAPreverse}). The calculated time-dependent eigenvalue ${E\left(t\right) \mathord{\left/{\vphantom{E\left(t\right) \left(2\pi \hbar \right)}}\right.\kern-\nulldelimiterspace} \left(2\pi \hbar \right)} $ of the Hamiltonian from equation~\ref{H_2STIRAP}, which corresponds to the initial state  $\ket{g}$, is shown in figure~\ref{STIRAPreverse}(c) for constant detuning $\delta/(2\pi)=10$~MHz and in figure~\ref{STIRAPreverse}(d) for $\delta/(2\pi)=10\,\mathrm{MHz}\times \mathrm{sgn}(t)$. Figures~\ref{STIRAPreverse}(e) and \ref{STIRAPreverse}(f) show comparison of the numerically calculated phase of the ground state $\ket{g}$ with $\mathrm{Arg}\left[\exp \left[-\frac{i}{\hbar } \int _{-T}^{t}E\left(t'\right)dt' \right]\right]$. Good agreement is observed when the ground-state population, shown in figures~\ref{STIRAPreverse}(g) and \ref{STIRAPreverse}(h), respectively, is non-zero. The total area in figure~\ref{STIRAPreverse}(d) $\int _{-T}^{T}E\left(t'\right)dt'=0$. This explains the compensation of the dynamical phase by switching the sign of the detuning during STIRAP in the regime of Rydberg blockade.

\begin{figure}[!t]
\includegraphics[width=\columnwidth]{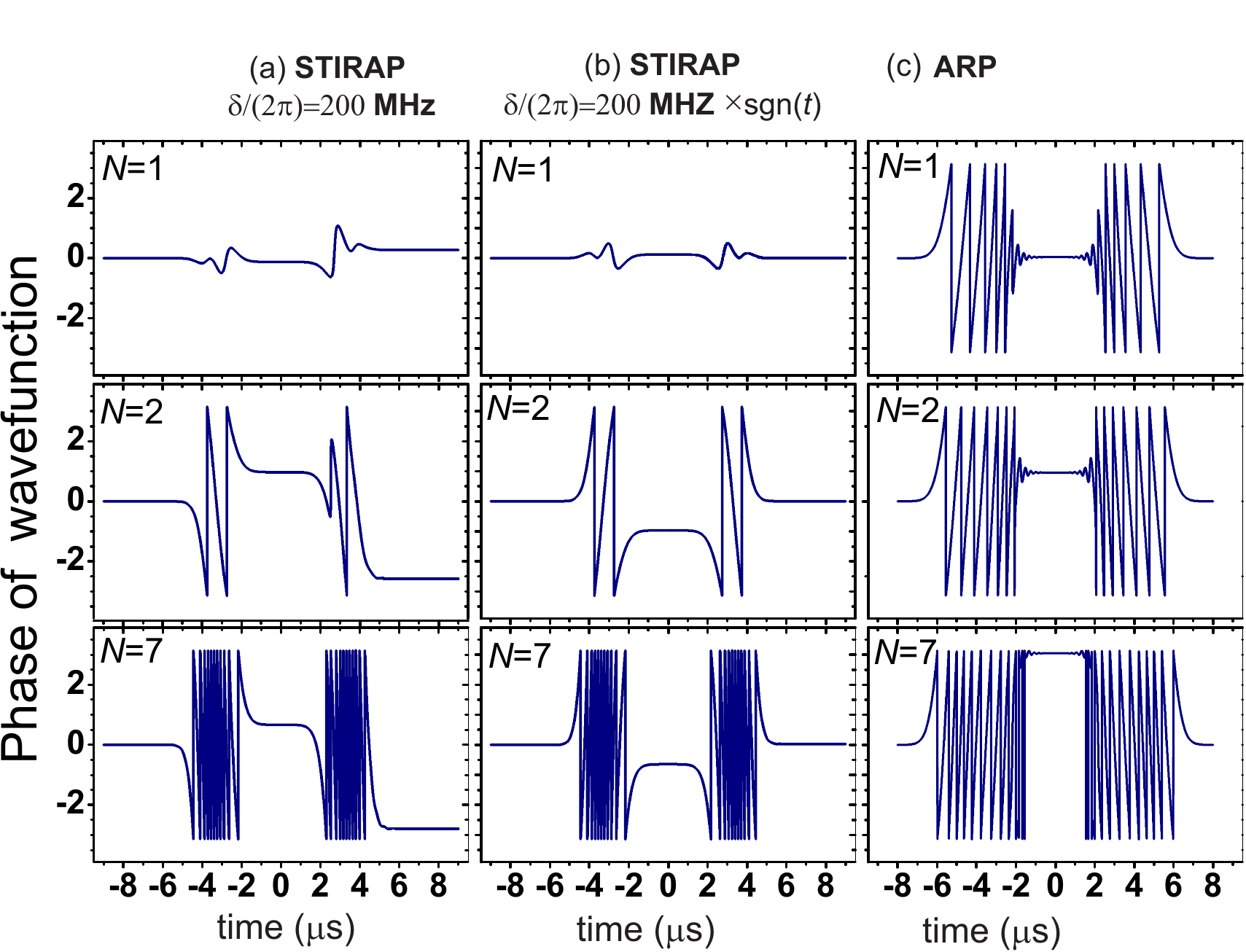}
\vspace{-.5cm}
\caption{
\label{DoubleSequence}(Color online).
The calculated time dependence of the phase 
of the collective ground state amplitude for $N=1,2,7$ atoms (top to bottom) for double STIRAP sequence (a)  with  $\delta/2\pi = 200~\mathrm{MHz}$, (b) with  $\delta/2\pi = 200~\mathrm{MHz} \times \mathrm{sgn}\left( t \right)$, and (b) for a double ARP pulse sequence with phase inversion. The STIRAP sequence used Gaussian pulses described by equation~\ref{STIRAP_Rabi} with $t_1 =-1\,\mu$s, $t_{2} =1\,\mu$s, $w=1\,\mu$s  $\Omega_{P0} /\left(2\pi\right)=40$~MHz, $\Omega_{S0} /\left(2\pi\right)=30$~MHz and detuning from the intermediate state $\delta /\left(2\pi\right)=200$~MHz. The single ARP pulse used $\Omega_0(t)=\Omega_{0} e^{-t^2/2w^2}$ with $\Omega_0/2\pi=2~\rm MHz$, $w=1~\mu\rm s$, and linear chirp rate $\alpha/(2 \pi)=1 ~{\rm MHz}/\mu\rm s$.
}
\end{figure}

Similar time dependence of the phase of the ground state was observed in our calculations for mesoscopic ensembles with arbitrary number of atoms. The STIRAP sequence used Gaussian pulses described by equation~(\ref{STIRAP_Rabi}) with $t_1 =-1\,\mu$s, $t_{2} =1\,\mu$s, $w=1\,\mu$s,  $\Omega_{P0} /\left(2\pi\right)=40$~MHz, $\Omega_{S0} /\left(2\pi\right)=30$~MHz and detuning from the intermediate state $\delta /\left(2\pi\right)=200$~MHz. The single ARP pulse used $\Omega_0(t)=\Omega_{0} e^{-t^2/2w^2}$ with $\Omega_0/2\pi=2~\rm MHz$, $w=1~\mu\rm s$, and linear chirp $1 ~{\rm MHz}/\mu\rm s$~\cite{Beterov2011}. The phase accumulation of the atomic wavefunction can be compensated by switching the sign of the detuning between two STIRAP pulses, or by switching the phase between two ARP pulses, as shown in figure~\ref{DoubleSequence}(a). For a double STIRAP sequence with the same detuning throughout the accumulated phase depends on $N$ [figure~\ref{DoubleSequence}(c)], while the phase change is zero, independent of $N$, when we switch the sign of detuning $\delta$ between the two STIRAP sequences
[figure~\ref{DoubleSequence}(d)]. A similar phase cancellation  occurs for $\pi$ phase shifted  ARP pulses
[figure~\ref{DoubleSequence}(e)], which can be implemented using an acousto-optic modulator.

\section{Quantum gates based on adiabatic passage in mesoscopic ensembles}

The double STIRAP and ARP sequences can be used for implementation of single-qubit and two-qubit gates with mesoscopic atomic ensembles used as qubits. A qubit can be encoded in an $N$
atom ensemble with the logical states~\cite{Lukin2001} 
\bea
\ket{\bar 0} = \ket{000...000}, \\\nonumber
\ket{\bar 1}' =\frac{1}{\sqrt{N}}\sum_{j=1}^N \ket{000 ... 1_j ...000}. 
\eea

\noindent Levels $\ket{0},\ket{1}$ are atomic hyperfine ground states.
Coupling between these states is mediated by the singly excited Rydberg state 
\be
\ket{\bar r}' =\frac{1}{\sqrt{N}}\sum_{j=1}^N \ket{000 ... r_j ...000}.
\ee
\noindent
Rydberg blockade only allows single excitation of  $\ket{r}$
so the states $\ket{\bar 0}$ and $\ket{\bar r }'$ experience a collectively enhanced coupling rate $\Omega_N=\sqrt
N\Omega$. States $\ket{\bar r}'$ and $\ket{\bar 1 }'$ are  coupled at the single atom rate $\Omega$. State $\ket{\bar
1}'$ is produced by the sequential application of $\pi$ pulses $\ket{\bar 0}\rightarrow\ket{\bar r}'$ and $\ket{\bar
r}'\rightarrow\ket{\bar 1}'$. However, adiabatic passage cannot be used for arbitrary rotations of the quantum states on the Bloch sphere. Therefore more complex schemes of quantum gates are required.

\begin{center}
\begin{figure}[!t]
 \center
\includegraphics[width=\columnwidth]{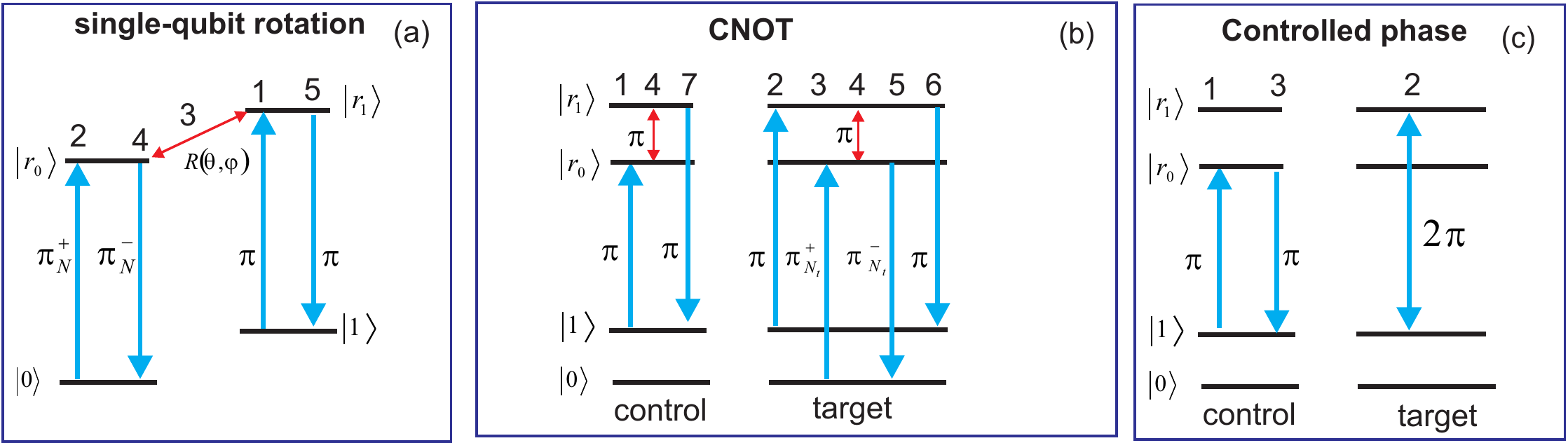}
\vspace{-.5cm}
\caption{
\label{EnsembleGates}(Color online).
Schemes of the quantum gates based on phase-preserving adiabatic passage: (a) single-qubit rotation with a mesoscopic atomic ensemble; (b) CNOT gate with two mesoscopic atomic ensembles used as qubits; (c) CZ gate with two atoms using adiabatic passage of Stark-tuned F\"{o}rster resonance; (d) ) CNOT gate with two atoms using adiabatic passage of Stark-tuned F\"{o}rster resonance.
}
\end{figure}
\end{center}

\noindent
Our proposal is shown in figure~\ref{EnsembleGates}(a)-(c), where we use two hyperfine sublevels of the ground state of alkali-metal atom $\ket{0},\ket{1}$ for storage of quantum information and two $\ket{r_0},\ket{r_1}$ auxiliary Rydberg states coupled by the microwave radiation for coherent rotation of the ensemble qubit on arbitrary angles after Rydberg excitation. 

\textit{Single-qubit gates}

We define the ensemble states as:
\bea
\ket{\bar 0} &=&\ket{000...000}\\\nonumber
\ket{\bar 1}' &=& \frac{1}{\sqrt{N}}\sum_{j=1}^N \ket{000 ... 1_j ...000}\\\nonumber
\ket{\bar r_0}' &=& \frac{1}{\sqrt{N}}\sum_{j=1}^N \ket{000 ... (r_0)_j ...000}\\\nonumber
\ket{\bar r_1}' &=& \frac{1}{\sqrt{N}}\sum_{j=1}^N \ket{000 ... (r_1)_j ...000}.
\eea

\noindent The basic idea of our gate, shown in figure~\ref{EnsembleGates}(a) is to transfer of the population of the initially excited Rydberg state $\ket{r_0}$ to an auxiliary Rydberg level $\ket{r_1}$, which can be done by coherent Rabi pulse, creating the superposition of two collective states, each of them having a single Rydberg excitation. Due to Rydberg blockade the second STIRAP pulse will transfer the collective state $\ket{\bar r_0}'$ back to the state $\ket{\bar 0}$ while the state $\ket{\bar r_1}'$ will remain unchanged due to the presence of a single Rydberg excitation in the state $\ket{r_1}$ which blocks the transition $\ket{0} \rightarrow \ket{r_0}$. After the end of the second STIRAP sequence the state $\ket{r_1}$ is transfered to the state $\ket{1}$ by a single $\pi$ pulse.

Pulse areas independent of $N$ on the $\ket{0} \leftrightarrow \ket{r_0}'$ transition can be implemented with STIRAP or ARP
as described above.
We will define the logical basis states as $\ket{\bar 0} = \ket{000...000},$
$\ket{\bar 1} = e^{\imath\chi_N} \ket{\bar 1}',$ and $\ket{\bar r} = e^{\imath\chi_N}\ket{\bar r}'.$ Here $\chi_N$ is
the phase produced by a single $N$-atom STIRAP pulse with positive detuning.  We assume that we do not know the value
of $N$, which may vary from qubit to qubit,  and therefore $\chi_N$ is also unknown, but has a definite value for fixed
$N$.
The logical states are $ \ket{\bar 0}$ and $\ket{\bar 1} = e^{\imath\chi_N} \ket{\bar 1}'$. The auxiliary Rydberg states are defined as 
\bea
\ket{\bar r_0} &=&e^{\imath\chi_N}\ket{\bar r_0}' \\\nonumber
\ket{\bar r_1} &=& e^{\imath\chi_N}\ket{\bar r_1}'.
\eea 
\noindent Starting with a qubit state $\ket{\psi}=a\ket{\bar 0} + b \ket{\bar 1}$
we perform a sequence of pulses 1-5, shown in figure~\ref{EnsembleGates}(a), giving 
the sequence of states 
\bea
\ket{\psi_1}&=&a\ket{\bar 0} + ib \ket{\bar r_1}\nonumber\\
\ket{\psi_2}&=&a \ket{\bar r_0} + ib \ket{\bar r_1}\nonumber\\
\ket{\psi_3}&=&a' \ket{\bar r_0} - ib' \ket{\bar r_1}\label{eq.1qubit}\\
\ket{\psi_4}&=&a' \ket{\bar 0} - ib' \ket{\bar r_1}\nonumber\\
\ket{\psi_5}&=&a' \ket{\bar 0} +b' \ket{\bar 1}.\nonumber
\eea

\noindent
The final state $\ket{\psi}=a'\ket{\bar 0} + b' \ket{\bar 1}$ is arbitrary and is selected by the rotation $R(\theta,\phi)$,  in step 3:
$\left({{\begin{array}{*{20}c}
a'\\-b'\\\end{array}}}\right)={\bf R}(\theta,\phi)
\left({{\begin{array}{*{20}c}a\\b\\\end{array}}}\right)$

\textit{CNOT:}
The proposed scheme is an extension of the experiment~\cite{Isenhower2010} and modification of our previous proposal~\cite{Beterov2013}.
Starting with an arbitrary two-qubit state
$\ket{\psi}=a\ket{\bar 0\bar 0} + b \ket{\bar 0 \bar 1} + c\ket{ \bar 1 \bar 0} + d \ket{\bar 1 \bar 1}$
we generate the sequence of states
\bea
\ket{\psi_1}&=&a \ket{{\bar 0}{\bar 0}} + b \ket{{\bar 0} {\bar 1}} +i c\ket{ {\bar r_0}{ \bar 0}} + i d \ket{{\bar r_0} {\bar 1}}\nonumber\\
\ket{\psi_2}&=&a \ket{{\bar 0}{\bar 0}} +i b \ket{{\bar 0} {\bar r_1}} +i c\ket{ {\bar r_0}{ \bar 0}} + i d \ket{{\bar r_0} {\bar 1}}\nonumber\\
\ket{\psi_3}&=&a \ket{{\bar 0}{\bar r_0}} +i b \ket{{\bar 0} {\bar r_1}} + i c\ket{ {\bar r_0}{ \bar 0}} + i d \ket{{\bar r_0} {\bar 1}}\nonumber\\
\ket{\psi_4}&=&i a \ket{{\bar 0}{\bar r_1}}  - b \ket{{\bar 0} {\bar r_0}} - c\ket{ {\bar r_1}{ \bar 0}} - d \ket{{\bar r_1} {\bar 1}}\label{eq.2qubit}\\
\ket{\psi_5}&=&i a \ket{{\bar 0}{\bar r_1}} - b \ket{{\bar 0} {\bar 0}} - c\ket{ {\bar r_1}{ \bar 0}} - d \ket{{\bar r_1} {\bar 1}}
\nonumber\\
\ket{\psi_6}&=& -a \ket{{\bar 0}{\bar 1}} -b \ket{{\bar 0} {\bar 0}} -c\ket{ {\bar r_1}{ \bar 0}} - d \ket{{\bar r_1} {\bar 1}}
\nonumber\\
\ket{\psi_7}&=&-a \ket{{\bar 0}{\bar 1}}  - b \ket{{\bar 0} {\bar 0}} -  i c\ket{ {\bar 1}{ \bar 0}} - i d \ket{{\bar 1} {\bar 1}}
\nonumber.
\eea

\noindent The gate matrix is therefore 
\begin{equation}
U_{CNOT} = \left( {{\begin{array}{*{20}c}
0 & -1 & 0 & 0 \\
-1  & 0 & 0 & 0 \\\label{CNOT}
0  & 0  & -i & 0 \\
0  &  0 & 0 & -i \\
\end{array}} } \right).
\end{equation}
\noindent which can be converted into a standard CNOT gate with a single qubit rotation. 

\textit{Controlled phase gate}

The controlled phase gate is implemented in the way similar to CNOT with replacement of the amplitude-swap sequence by controlled $2\pi$ rotation of the target qubit which 
could be switched on and off by excitation of the control qubit into the Rydberg state. 

We find that arbitrary single qubit rotations in the basis $\ket{\bar 0},\ket{\bar 1}$ can be performed with high
fidelity, without precise knowledge of $N$, by accessing several Rydberg levels $\ket{r_0}$, $\ket{r_1}$ as shown in
figure~\ref{EnsembleGates}(a).  Depending on the choice of
implementation, to be discussed below,  this may be given by a one- or two-photon microwave pulse, with Rabi frequency $\Omega_3$.
Provided  states $\ket{r_0}, \ket{r_1}$ are strongly interacting, and
limit the number of excitations in the  ensemble to one, the indicated sequence is obtained.  

The five pulse sequence we describe here is more complicated than the three pulses needed for an arbitrary  single
qubit gate in the approach of Ref.~\cite{Lukin2001}. The reason for this added complexity is that the special phase
preserving property of the double STIRAP or ARP sequences requires that all population is initially in one of the
states connected by the  pulses. The sequence of pulses in figure~\ref{EnsembleGates}(a) ensures that this condition is
always satisfied.

All pulses except number 4 in the CNOT
sequence are optical and are localized to either the control or target qubit. Pulse 4 is a microwave field and drives a
$\pi$ rotation on both qubits. As for the single qubit gate the requirement for high fidelity operation is that the
interactions $\ket{r_0} \leftrightarrow \ket{r_0}$, $\ket{r_1} \leftrightarrow \ket{r_1}$, $\ket{r_0} \leftrightarrow
\ket{r_1}$ all lead to full blockade of the ensembles. Since the frequency of pulse 4, which is determined by the energy separation of states
$\ket{r_0},\ket{r_1}$, can be chosen to be very different from the qubit frequency given by the energy separation of
states $\ket{0},\ket{1}$ the application of microwave pulses will not lead to crosstalk in an array of ensemble qubits.

\section{Adiabatic passage across Stark-tuned F\"{o}rster resonance} 

\subsection{Scheme of two-qubit gates}

Another than Rydberg blockade approach to build a two-qubit gate is based on controlled phase shifts of collective states of two qubits due to interaction between Rydberg atoms~\cite{Ryabtsev2005}. The interaction strength should be adjusted to provide a certain phase shift (for example $\pi$),  during the interaction time.  

\begin{center}
\begin{figure}[!t]
 \center
\includegraphics[width=\columnwidth]{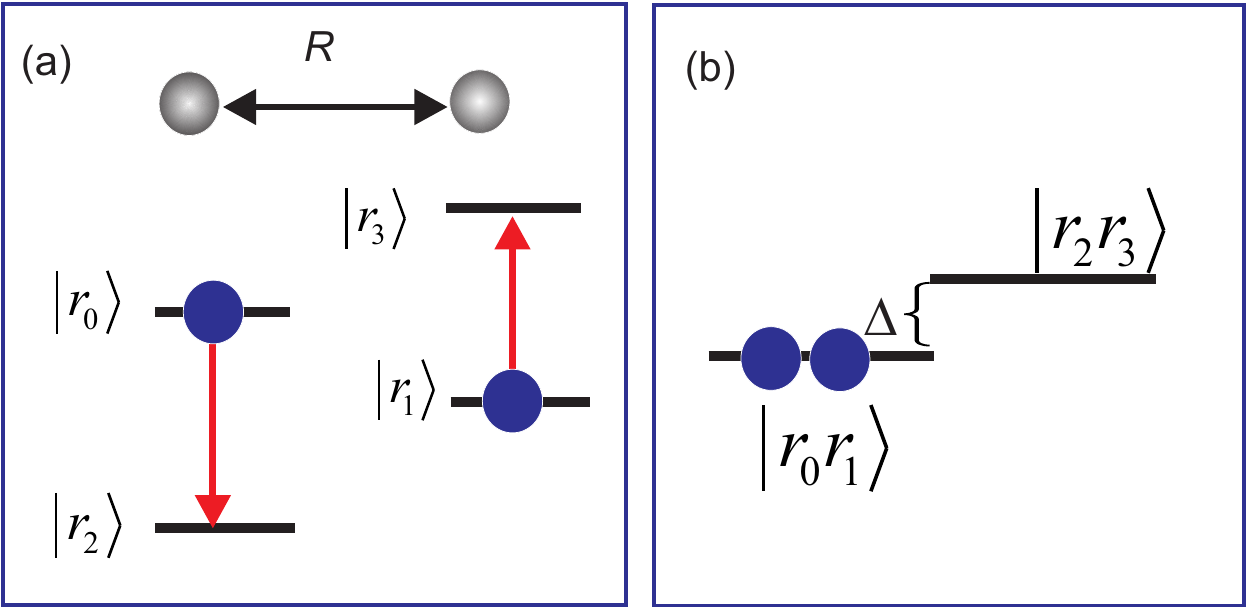}
\vspace{-.5cm}
\caption{
\label{Forster}(Color online).
 (a) Scheme of the F\"{o}rster resonant energy transfer between two Rydberg atoms;  (b) Energy defect of the F\"{o}rster  resonance is the difference between the energies of the final and initial collective states of two interacting atoms.
}
\end{figure}
\end{center}

When two atoms are excited into Rydberg states $\ket{r_0}$ and  $\ket{r_1}$, which can be either different or identical, the dipole-dipole interaction may result to transitions to neighboring states $\ket{r_2}$ and  $\ket{r_3}$, as shown in figure~\ref{Forster}(a). This energy exchange between interacting atoms is an analog of the F\"{o}rster resonant energy transfer. 

The detuning from the F\"{o}rster  resonance  $\hbar\Delta= U(\ket{r_2 r_3})-U(\ket{r_0 r_1})$ is  a difference between the energies of the final $\ket{r_2r_3}$  and initial  $\ket{r_0r_1}$ collective states of the two-atom system. Due to the difference in the polarizabilities of initial and final states, in some cases it is possible to tune the F\"{o}rster   energy defect in external electric field (Stark tuned F\"{o}rster   resonance)~\cite{Safinya1981}.

 Stark-tuned F\"{o}rster resonances for two Rydberg atoms were first reported in Ref.~\cite{Ryabtsev2010}. The rf-assisted "inaccessible" Stark-tuned F\"{o}rster resonances have been demonstrated in Ref.~\cite{Tretyakov2014}. 

If two Rydberg atoms are frozen in space, dipole-dipole interaction at a F\"{o}rster resonance induces the Rabi-like coherent population oscillations between collective states of these atoms~\cite{Ryabtsev2010a}. Such oscillations have been demonstrated recently for two Rb Rydberg atoms in two optical dipole traps~\cite{Ravets2014}. The frequency of these collective oscillations is sensitive to variations of the interaction energy due to fluctuations of the spatial uncertainty of the atoms within the optical dipole traps. For example, a 10\% variation of the distance between the trapped atoms results in a 25\% variation of the interaction energy due to the $1/R^{3}$ dependence of the energy of dipole-dipole interaction on distance $R$ between the atoms. This can substantially increase the phase gate error. We propose to overcome this difficulty by using a double adiabatic rapid passage across Stark-tuned F\"orster resonances with a deterministic phase accumulation. This technique is similar to Stark-chirped rapid adiabatic passage, which is based on a laser-induced Stark shift~\cite{Shore2011, Yatsenko2002}. Landau-Zener control of the Stark-tuned F\"{o}rster resonances can be used for implementation of two-qubit gates~\cite{Huang2018}. 

\begin{center}
\begin{figure}[!t]
 \center
\includegraphics[width=\columnwidth]{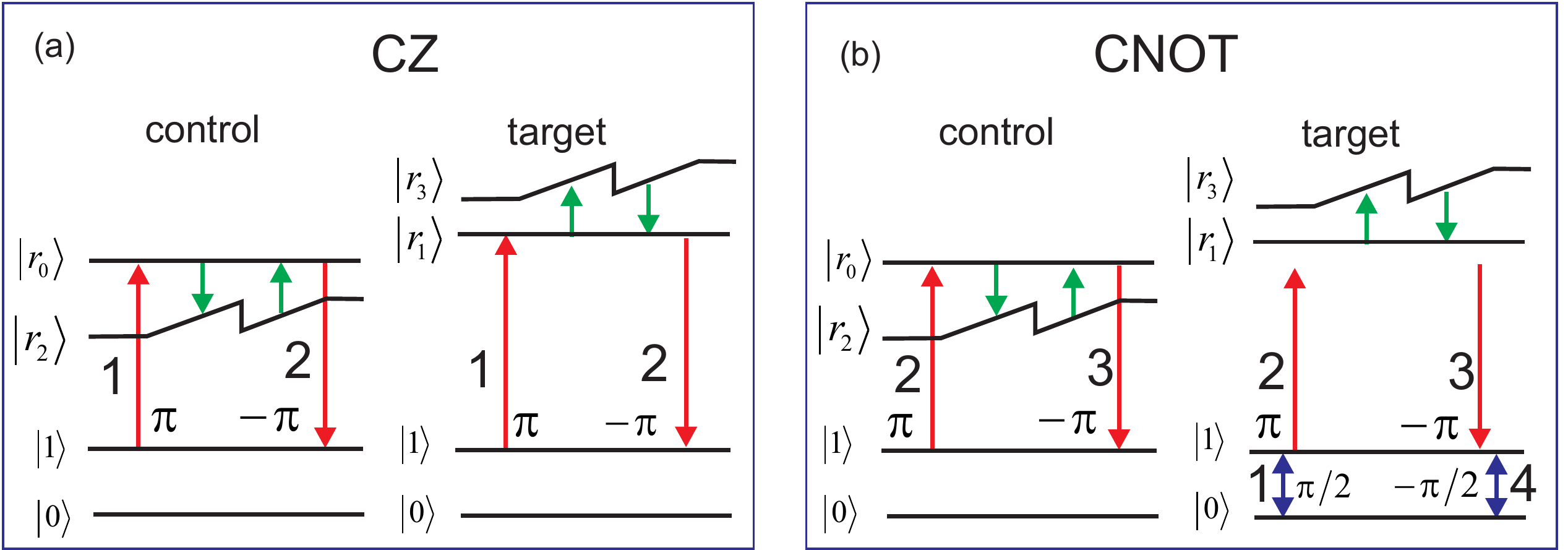}
\vspace{-.5cm}
\caption{
\label{ForsterGates}(Color online).
 (a) Scheme of a CZ gate using double adiabatic rapid passage across Stark-tuned F\"{o}rster resonance. Two atoms are excited to  Rydberg states. An external electric field shifts the energy levels of the Rydberg atoms so that the F\"orster resonance  is passed adiabatically two times. Then the atoms are de-excited to ground state. The phase shift is deterministically accumulated if both atoms are initially prepared in state $\ket{1}$; (b) Scheme of a CNOT gate. Two additional $\pi/2$ pulses rotate the target qubit around the \textit{y} axis in the opposite directions.
}
\end{figure}
\end{center}

A scheme of controlled-Z gate is shown in Fig.~\ref{ForsterGates}(a). Two optical dipole traps with one atom in each trap are located at a distance $R$ between them. The two atoms are simultaneously excited to Rydberg state $\ket{r}$ by a $\pi$ laser pulse labeled as 1. The distance between the traps must be sufficiently large to avoid the effect of Rydberg blockade~\cite{Lukin2001}. A time-dependent external electric field shifts the collective energy levels so that the F\"orster resonance $\ket{r_0r_1} \to \ket{r_2r_3}$ is passed adiabatically two times. This results in a deterministic phase shift of state $\ket{rr}$. After the end of adiabatic passage the atoms are de-excited to ground state by a 3$\pi$ laser pulse labeled as 2. 

The phase shift due to Rydberg-Rydberg interaction is accumulated only in the case when both atoms are initially prepared in state 
$\ket{1}$ and then excited to Rydberg states $\ket{r_0}$ and $\ket{r_1}$  . If one of the atoms (or both of them) is initially in the state $\ket{0}$, no phase shift occurs. 

\subsection{Adiabatic Rapid Passage with nonlinear detuning}

The energy of dipole-dipole interaction of two Rydberg atoms is determined by the interatomic distance and cannot be changed on short timescales. Therefore we need to consider the adiabatic rapid passage with constant Rabi frequency. To achieve high fidelity of the population transfer, we use a nonlinear time dependence of the detuning from the resonance: 
\begin{center}
\begin{figure}[!t]
 \center
\includegraphics[width=\columnwidth]{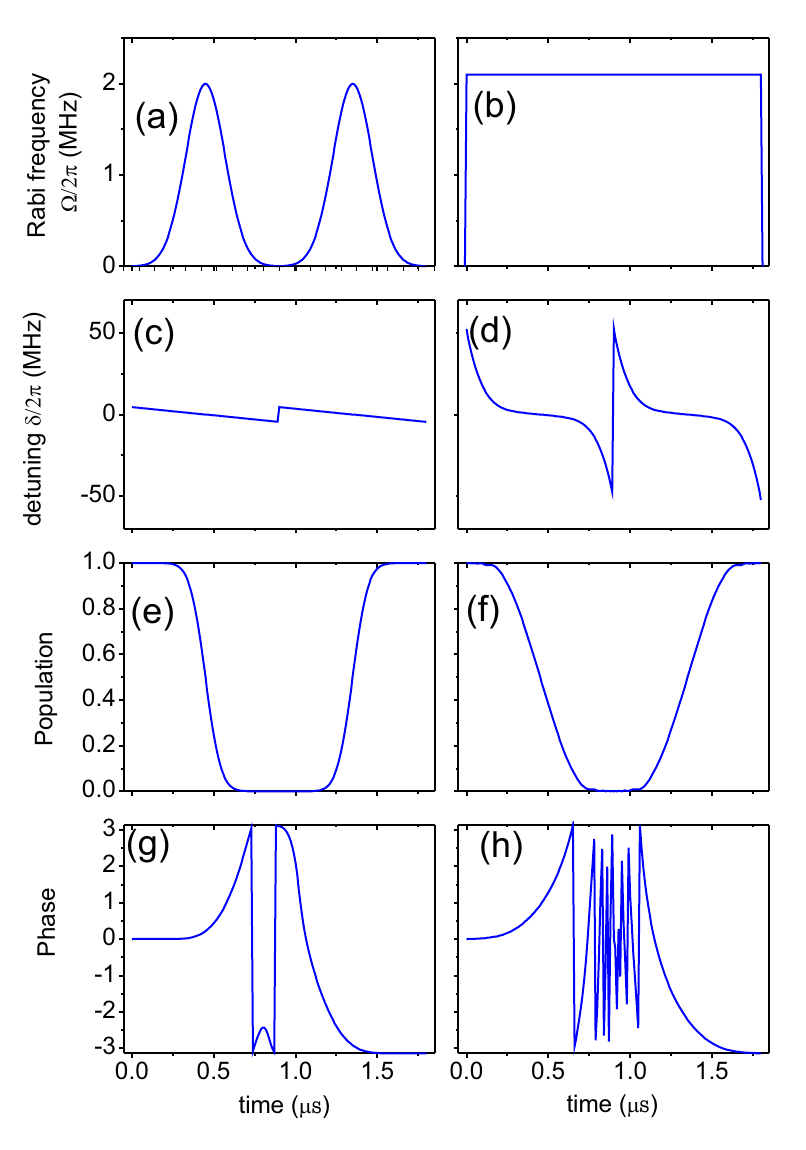}
\vspace{-.5cm}
\caption{
\label{Nonlin}(Color online) Comparison between the schemes of double adiabatic rapid passage with linearly chirped Gaussian pulses (left-hand panel), and with constant Rabi frequency and nonlinear time dependence of detuning  from the resonance (right-hand panel). (a),(b) Time dependence of Rabi frequency $\Omega_0 \left(t\right)$; (c),(d) Time dependence of detuning from the resonance $\delta \left(t\right)$; (e),(f) Time dependence of the population of state $\ket{1} $; (g),(h) Time dependence of the phase of state $\ket{1}  $.
}
\end{figure}
\end{center}

\be
\label{NonlinDelta}
\delta _{j} \left(t\right)=s_{1} \left(t-t_{j} \right)+s_{2} \left(t-t_{j} \right)^{3}.  
\ee

\noindent The detuning is slowly varied across the resonance and is rapidly increased before and after the resonance. Figure~\ref{Nonlin} illustrates the difference between the conventional scheme of adiabatic rapid passage, which uses chirped Gaussian pulses with linear time dependence of detuning (left-hand panel), and the scheme of adiabatic rapid passage with constant Rabi frequency and nonlinear time dependence of detuning (right-hand panel). The parameters of the pulses for the left panel of figure~\ref{Nonlin} are $\Omega_0/2\pi=10$~MHz, $w=0.12\; \mu s$, and $\delta_j \left(t\right)=s_1 \left(t-t_{j} \right)$ with 
$s_1=-100\;\mathrm{MHz}/\mu s$. The parameters of the pulses for the right-hand panel of figure~\ref{Nonlin} are $\Omega_0/2\pi=2.1$~MHz, 
$s_1/2\pi=-10\;\mathrm{MHz}/\mu s$, and $s_2/2\pi=-2000\;\mathrm{MHz}/\mu s^3$. The centers of the pulses are located at times $t_1 =0.5\; \mu s$ and $t_{2} =1.5\; \mu s$. The population error for the final state of the system is found to be below $4\times 10^{-5}$ in both cases. The phase shift is equal to $\pi$ in both cases.

\subsection{Stark-tuned adiabatic rapid passage}

Stark-tuned F\"{o}rster resonance required for the implementation of the proposed scheme must meet the following criteria: (i) the lifetimes of Rydberg states must be sufficiently long to avoid the decay of coherence during the gate operation due to spontaneous and BBR-induced transitions; (ii) initial F\"{o}rster energy defect must be sufficiently large to allow for rapid turning off the interaction between atoms at the beginning and the end of the adiabatic passage; (iii) selected interaction channel must be well isolated from the other channels to avoid break-up or dephasing of the adiabatic population transfer.

\begin{center}
\begin{figure}[!t]
 \center
\includegraphics[width=\columnwidth]{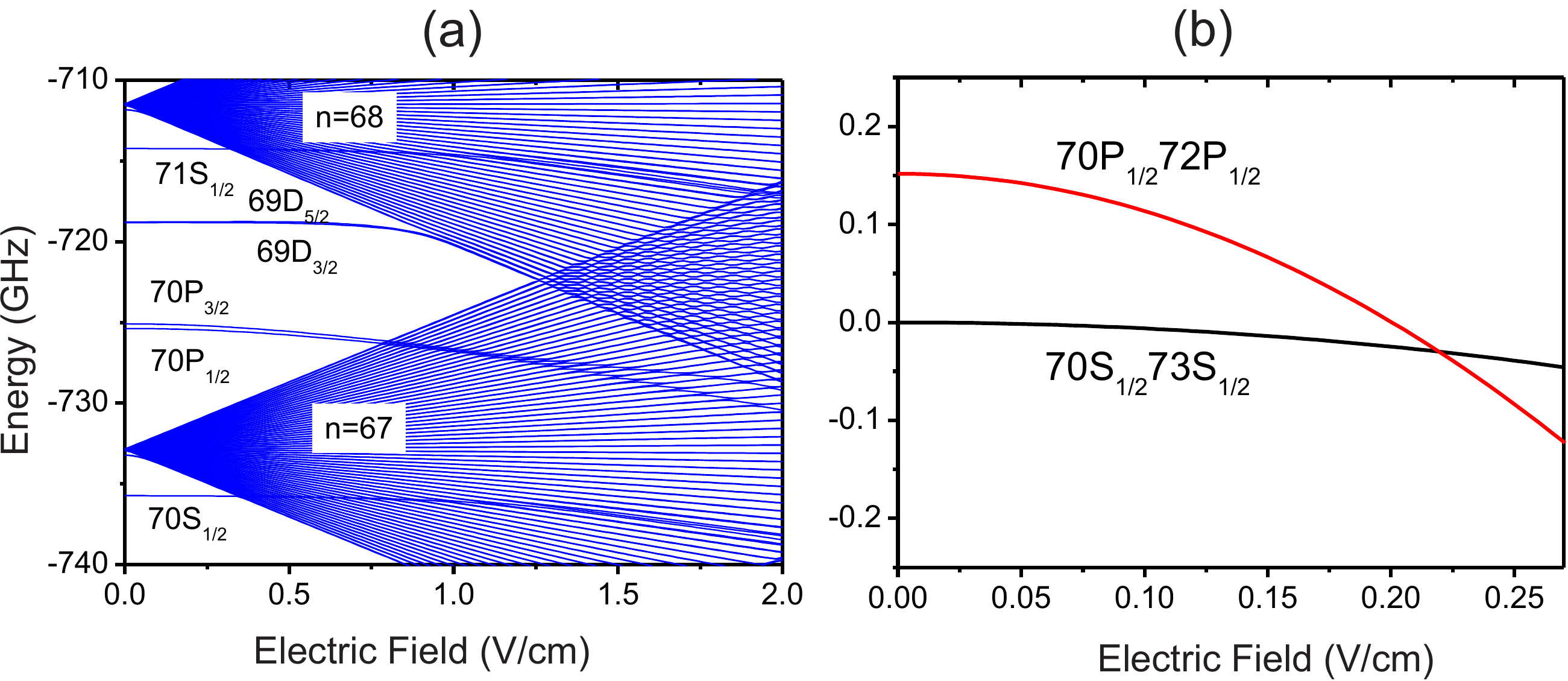}
\vspace{-.5cm}
\caption{
\label{Stark}(Color online)  (a) Stark diagram for Rb Rydberg states with $\left|m_{j} \right|=1/2 $; (b) Stark diagram for the collective two-atom states $\ket{ 70S,73S}$ and $\ket{70P,72P} $  in Rb.  The F\"{o}rster resonance occurs in the electric field $E=0.222$~V/cm.
}
\end{figure}
\end{center}

In our previous work~\cite{Beterov2015} we have studied the structure of the F\"{o}rster resonances $\ket{nS,n'S} \to \ket{nP,\left(n'-1\right)P}$ in Rb and Cs Rydberg atoms. We have selected the $\ket{70S_{1/2} ,73S_{1/2}} \to \ket{70P_{1/2},72P_{1/2}}$ Stark-tuned F\"{o}rster resonance in Rb for the further numerical simulations. This resonance has the energy defect  $\Delta/(2 \pi) =152$~MHz in a zero electric field. In contrast to the resonances involving $\ket{nP_{3/2}}$ states, this resonance has no Stark splitting in the electric field. The Stark diagram for Rb Rydberg states with $\left|m_j \right|=1/2$ is shown in figure~\ref{Stark}(a). The dc electric field is aligned along the \textit{z} axis. The exact F\"{o}rster resonance occurs in the electric field $E=0.222$~V/cm, as shown in figure~\ref{Stark}(b).

\begin{center}
\begin{figure}[!t]
 \center
\includegraphics[width=\columnwidth]{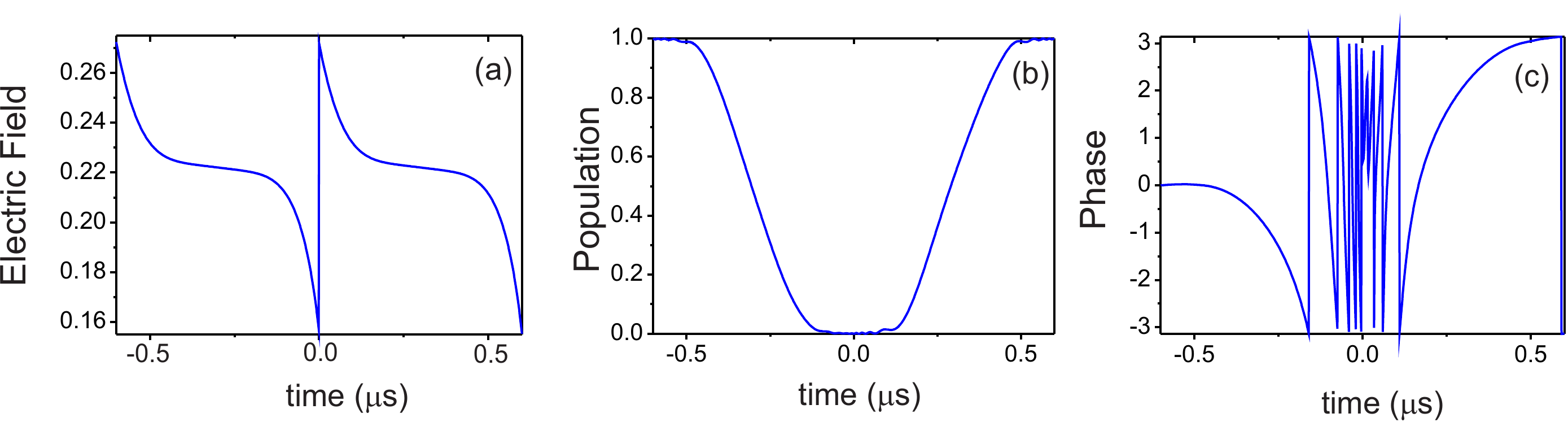}
\vspace{-.5cm}
\caption{
\label{CZ}(Color online) Time dependence of the electric field required for double adiabatic passage of the  Stark-tuned F\"{o}rster resonance $\ket{70S_{1/2} ,73S_{1/2}}\to\ket{70P_{1/2} ,72P_{1/2}}$ in Rb Rydberg atoms; (b) Time dependence of population of the collective state $\ket{70S_{1/2} ,73S_{1/2}}$; (c) Time dependence of the phase of the collective state $\ket{70S_{1/2} ,73S_{1/2}}$.
}
\end{figure}
\end{center}

The time dependence of the electric field required to form the nonlinearly shaped detuning $\delta _{j} \left(t\right)=s_{1} \left(t-t_{j} \right)+s_{2} \left(t-t_{j} \right)^5$ of the $\ket{70S_{1/2} ,73S_{1/2}} \to \ket{70P_{1/2} ,72P_{1/2}}$ F\"{o}rster resonance with $s_1/2 \pi= 22.6\;\mathrm{MHz} / \mu s$ and $s_2 /2\pi =28800\; \mathrm{MHz} /\mu s^{5}$, $t_1 =-0.3\,\mu$s and $t_2 =0.2993\,\mu$s is shown in Fig.~\ref{CZ}(a). The value of $t_2$ was selected to reduce the influence of off-resonant channels of F\"{o}rster interaction.
The time dependence of the population [figure~\ref{CZ}(b)] and phase [figure~\ref{CZ}(c)] of the collective $\ket{ 70S_{1/2} ,73S_{1/2}} $ state for two interacting Rydberg atoms located at distance $R$=15.5~$\mu $m along the \textit{z} axis was calculated taking into account Stark sublevels of Rydberg states.  Our calculations have shown that the variation of the interatomic distance leads to small phase changes at the end of the adiabatic passage, thus evidencing that our method to perform two-qubit quantum gates is insensitive to the atom position uncertainty. 

\section{Conclusion}
This tutorial is a brief review of the methods of control of populations and phases of the collective states in ensembles of interacting atoms using adiabatic passage. 
Adiabatic Rapid Passage and Stimulated Raman Adiabatic Passage are commonly used for population inversion in atomic and molecular systems due to their reduced sensitivity to fluctuations of the parameters of the experiment. We have shown that similar advantages can be demonstrated for mesoscopic ensembles of strongly interacting atoms. We have studied the phase dynamics of the collective states of atomic ensembles and demonstrated the ability for their precise control, as required for quantum information processing. Our results are confirmed both by numeric simulations and by analytic formulas in the adiabatic approximation. Adiabatic passage of the Stark-tuned F\"{o}rster resonance is an interesting technique which can be used for implementation of controlled phase gates based on long-range Rydberg interactions. Several schemes of quantum gates for atomic ensembles and single atoms have been developed. The experimental implementation of the proposed methods can be useful to reduce the sensitivity of the gate fidelities to fluctuations of the experimental parameters (number of atoms in the trap, interatomic distances, etc.).

\section*{Acknowledgements}
This work was supported by the Russian Science Foundation Grant No. 18-12-00313 (for quantum gates) and Russian Foundation For Basic Research Grant No.17-02-00987 (for numerical simulations).

\section*{References}


%

\end{document}